\documentclass[useAMS,usenatbib]{mn2e}

\usepackage{epsfig,color}
\usepackage{natbib}
\bibpunct{(}{)}{;}{a}{}{,}
\def\spose#1{\hbox to 0pt{#1\hss}}
\def\ltsimm{\mathrel{\spose{\lower 3pt\hbox{$\sim$}}
        \raise 2.0pt\hbox{$<$}}}
\def\gtsimm{\mathrel{\spose{\lower 3pt\hbox{$\sim$}}
        \raise 2.0pt\hbox{$>$}}}
\def\Mdot{\hbox{${\dot M}$}}

\def\km{{\rm\thinspace km}}
\def\cm{{\rm\thinspace cm}}

\def\s{{\rm\thinspace s}}
\def\yr{{\rm\thinspace yr}}
\def\g{{\rm\thinspace g}}
\def\kmps{\hbox{${\rm\km\s^{-1}\,}$}}

\def\erg{{\rm\thinspace erg}}

\def\Hz{{\rm\thinspace Hz}}

\def\ster{{\rm\thinspace ster}}
\def\ergps{\hbox{${\rm\erg\s^{-1}\,}$}}
\def\Rsol{\hbox{${\rm\thinspace R_{\odot}}$}}
\def\Msol{\hbox{${\rm\thinspace M_{\odot}}$}}

\def\Msolpyr{\hbox{${\rm\Msol\yr^{-1}\,}$}}
\def\pcm{\hbox{${\rm\cm^{-1}\,}$}}
\def\pcm2{\hbox{${\rm\cm^{-2}\,}$}}
\def\pcm3{\hbox{${\rm\cm^{-3}\,}$}}
\def\ergpscm3Hz{\hbox{${\rm\ergps\cm^{-3}\Hz^{-1}\,}$}}
\def\ergpscm3Hzster{\hbox{${\rm\ergps\cm^{-3}\Hz^{-1}\ster^{-1}\,}$}}
\def\gpcm3{\hbox{${\rm\g\cm^{-3}\,}$}}
\def\ergpcm2{\hbox{${\rm\erg\cm^{-2}\,}$}}
\def\ergpcm3{\hbox{${\rm\erg\cm^{-3}\,}$}}
\def\phpscm2{\hbox{${\rm photons\s^{-1}\cm^{-2}\,}$}}

\title[The 3D Hydrodynamics of Colliding Wind Binaries]{3D Models of Radiatively Driven Colliding Winds In Massive O+O Star Binaries: I. Hydrodynamics} 
\author[J.~M.~Pittard] {J. M. Pittard$^{1}$\thanks{E-mail:
jmp@ast.leeds.ac.uk}\\ $^{1}$School of Physics and Astronomy, The
University of Leeds, Leeds LS2 9JT, UK\\ }

\begin{document}

\date{Accepted ... Received ...; in original form ...}

\pagerange{\pageref{firstpage}--\pageref{lastpage}} \pubyear{2009}

\maketitle

\label{firstpage}

\begin{abstract}
The dynamics of the wind-wind collision in massive stellar binaries is
investigated using three-dimensional hydrodynamical models which
incorporate gravity, the driving of the winds, the orbital motion of
the stars, and radiative cooling of the shocked plasma. In this first
paper we restrict our study to main-sequence O+O binaries. The nature
of the wind-wind collision region is highly dependent on the degree of
cooling of the shocked plasma, and the ratio of the flow timescale of
the shocked plasma to the orbital timescale. The pre-shock wind speeds
are lower in close systems as the winds collide prior to their
acceleration to terminal speeds. Radiative inhibition may also reduce
the pre-shock wind speeds. Together, these effects can lead to rapid
cooling of the post-shock gas.  Radiative inhibition is less important
in wider systems, where the winds are accelerated to higher speeds
before they collide, and the resulting collision region can be largely
adiabatic.  In systems with eccentric orbits, cold gas formed during
periastron passage can persist even at apastron, before being ablated
and mixed into its surroundings and/or accelerated out of the system.
\end{abstract}


\section{Introduction}
\label{sec:intro}
The collision of the winds in massive star binary systems can produce
a rich array of phenomena.  Energetic particles are accelerated at the
wind-wind collision region \citep[WCR - see
e.g.][]{Eichler:1993,Dougherty:2003,Pittard:2006,Pittard:2006b},
producing non-thermal radio emission via the synchrotron process
\citep[see e.g.][]{Dougherty:2005}, and X-ray and $\gamma$-ray
emission via the inverse Compton and pion decay processes which should
extend to TeV energies \citep[]{Pittard:2006b,Reimer:2006}. Dust can
be formed either episodically producing transient infra-red outbursts
\citep{Williams:1996}, or continuously producing spiral-shaped
structures on the sky \citep[e.g.][]{Tuthill:1999,Marchenko:2002}.
Emission line profiles from the WCR show variable doppler shifts and
broadening as sight lines relative to the WCR change
\citep*[e.g.][]{Luhrs:1997,Henley:2003}.  Shock heating of the winds
produces bright thermal X-ray emission, which may exhibit signs of
non-equilibrium ionization in wide systems \citep{Pollock:2005}.

The hydrodynamics of colliding wind binaries (CWBs) have usually been
investigated using two-dimensional (axisymmetric) models with terminal
velocity winds \citep*[e.g.][]{Stevens:1992,Pittard:1997}. 3D
simulations which include orbital effects have been presented by
\citet{Walder:1998}, \citet{Pittard:1999}, and
\citet{Lemaster:2007}. In close binaries, the winds collide before
they reach terminal velocity, and their dynamics may be altered by the
companion star's radiation field \citep*{Stevens:1994,Gayley:1997,Gayley:1999}.
The driving of the winds and the dynamical effect of the companion's
gravitational and radiation fields has been included in 2D models of
V444~Cygni \citep{Gayley:1997}, $\iota$~Orionis \citep{Pittard:1998},
$\eta$~Carinae \citep{Pittard:1998b}, and Sanduleak~1
\citep{St-Louis:2005}. The only published 3D simulation with radiative
driving is of V444~Cygni \citep{Pittard:1999}. Other approaches based
on solutions to the ram-pressure balance have been presented by
\citet{Antokhin:2004} and \citet{Parkin:2008}.

In this work we examine the hydrodynamic properties of the wind-wind
collision in short period O+O binaries. We perform 3D calculations, so
that orbital effects can be included, and calculate the radiative
driving of the winds, so that the winds collide at realistic
velocities and some effects of the companion's radiation field can be
accounted for. In some of the simulations the winds collide at only a
fraction of their terminal velocities, and the lower postshock
temperatures and higher postshock densities which result enhance the
radiative cooling of the postshock gas.  We examine the dynamics of
the wind-wind collision in systems with circular and eccentric orbits,
and with equal and unequal winds. In Section~\ref{sec:setup} we
describe the details of our simulations and in
Section~\ref{sec:results} we present our results.
Section~\ref{sec:summary} summarizes and concludes this work.

\section{The Numerical Setup}
\label{sec:setup}
\subsection{Details of the hydrodynamics}
\label{sec:details_hydro}
Our three-dimensional simulations were conducted on a cartesian
grid using Eulerian hydrodynamics with piecewise parabolic interpolation
of the fluid variables. The code solves a Riemann problem at each zone
interface to determine the time-averaged values at the zone faces, and then
solves the equations of hydrodynamics:
\begin{eqnarray}
\label{eq:mass}
\frac{\partial \rho}{\partial t} + \nabla \cdot (\rho {\bf u}) & = & 0,\\  
\frac{\partial \rho {\bf u}}{\partial t} + \nabla \cdot (\rho {\bf u}u + P) & = & \rho {\bf f}, \\ 
\frac{\partial \rho \varepsilon}{\partial t} + \nabla \cdot [(\rho \varepsilon + P){\bf u}] & = & \left(\frac{\rho}{m_{\rm H}}\right)^{2}\Lambda(T) + \rho {\bf f} \cdot {\bf u}, 
\end{eqnarray} 
where $\varepsilon = {\bf u}^{2}/2 + e/\rho$ is the total specific
energy, $\rho$ is the mass density, $e$ is the internal energy
density, $P$ is the pressure, and $T$ is the temperature. We adopt an
ideal gas equation of state, $e = P/(\gamma - 1)$, with $\gamma=5/3$
as the ratio of specific heats. ${\bf f}$ is the force per unit
mass and includes gravity and radiative driving terms.  

The radiative cooling term, $\Lambda(T)$, is calculated from the MEKAL
thermal plasma code \citep{Mewe:1995,Kaastra:1992} distributed in
XSPEC (v11.2.0). The plasma is assumed to be in collisional ionization
equilibrium (but see Sec.~\ref{sec:postshock_eq}).  The temperature of
the pre-shock stellar winds is assumed to be maintained at $\approx
10^{4}\;$K through photoionization heating by the stars.  Gas in the
WCR which rapidly cools is prevented from cooling below this
temperature. Because of the high Mach numbers involved, the density
contrast of the hot plasma and cooled regions can be very high (the
density contrast across an isothermal shock of Mach number $M$ is
$\gamma M^{2}$). In reality, magnetic pressure may halt the
compression before this ratio is reached \citep[e.g.][]{Kashi:2007},
but this is an additional complication which is not considered
here. The code also contains several advected scalars which allow
tracking of which wind material is in which cell, and the ionization
age and temperature equilibration of the postshock gas (see
Section~\ref{sec:postshock_eq}).

The body forces acting on each hydrodynamic cell are the vector
summation of gravitational forces from each star, and continuum and
line driving forces from the stellar radiation fields.  The
computation of the line acceleration is based on a local
\citet{Sobolev:1960} treatment of the line transport, following the
standard \citet[][hereafter CAK]{Castor:1975} formalism developed for
single OB winds, augmented by the finite disk correction factor
developed by \citet*{Pauldrach:1986}. The vector radiative force per unit mass,
$\mathbf{g}^{\rm rad}$, is computed from an integral of the intensity
$I(\mathbf{\hat{n}})$ times the projected velocity gradient along
directional vectors $\mathbf{\hat{n}}$ within the solid angle covering
the stellar disk,
\begin{equation}
\label{eq:g_rad}
\mathbf{g}^{\rm rad} = \frac{\sigma_{e}^{1-\alpha} k}{c} \oint 
I(\mathbf{\hat{n}}) \left[\frac{\mathbf{\hat{n}} \cdot \nabla(\mathbf{\hat{n}} 
\cdot v)}{\rho v_{\rm th}}\right]^{\alpha} \mathbf{\hat{n}} d\mathbf{\Omega}.
\end{equation}
The integration over solid angle is performed with 8 directional
vectors.  $\alpha$ and $k$ are the standard CAK parameters,
$\sigma_{\rm e}$ is the specific electron opacity due to Thomson
scattering, and $v_{\rm th}$ is a fiducial thermal velocity calculated
for hydrogen\footnote{While we use the standard CAK $\alpha$ and
$k$ description, we note that the $k$ parameter does not represent a
physically meaningful quantity in its own right, and has an artifical
dependence on $v_{\rm th}$. A more meaningful description of the
strength of line driving was given by \citet{Gayley:1995}, who
introduced the $\bar{Q}$ parameter. $\alpha$ and $k$ values can easily
be converted into $\bar{Q}$ values, and are noted alongside the
$\alpha$ and $k$ values for each of our model stars in
Table~\ref{tab:stellar_params}.}. Shadowing by the companion star is
accounted for in our calculations, and in such cases only the visible
part of the stellar disk contributes to the radiative driving force.
The line driving is set to zero in cells with temperatures above
$10^{6}\;$K, since this plasma is mostly ionized. Further details
concerning the line force calculations can be found in
\citet{Gayley:1997}.

\begin{table}
\begin{center}
\caption[]{Assumed binary parameters for the models 
investigated. The semi-major axis is $34.26\;\Rsol$ in model cwb1,
$76.3\;\Rsol$ in models cwb2 and cwb3, and $55\;\Rsol$ in model cwb4.}
\label{tab:models}
\begin{tabular}{lllll}
\hline
\hline
Model & Stars & Period & Eccentricity & Wind mtm.\\
 & & (d) & (e) & ratio ($\eta$) \\
\hline
cwb1 & O6V+O6V & 3 & 0.0 & 1 \\
cwb2 & O6V+O6V & 10 & 0.0 & 1 \\
cwb3 & O6V+O8V & 10.74 & 0.0 & 0.4 \\
cwb4 & O6V+O6V & 6.1 & 0.$\overline{36}$ & 1 \\
\hline
\end{tabular}
\end{center}
\end{table}

\begin{table}
\begin{center}
\caption[]{Assumed stellar parameters for the models 
investigated. Note that the value of $\bar{Q}$ is fairly
constant, and shows less variation than $k$ in the two stellar
models constructed for this work.}
\label{tab:stellar_params}
\begin{tabular}{lllll}
\hline
\hline
Parameter/Star & O6V & O8V \\
\hline
Mass ($\Msol$) & 30 & 22 \\
Radius ($\Rsol$) & 10 & 8.5 \\
Effective temperature (K) & 38000 & 34000 \\ 
Mass-loss rate ($\Msolpyr$) & $2 \times 10^{-7}$ & $10^{-7}$ \\
Terminal wind speed ($\kmps$) & 2500 & 2000 \\
CAK $\alpha$ & 0.57 & 0.52 \\
CAK $k$ & 0.12 & 0.24 \\
$\bar{Q}$ & 258 & 308 \\
\hline
\end{tabular}
\end{center}
\end{table}

\begin{table}
\begin{center}
\caption[]{Details of the hydrodynamical grid of each model. The
length of the sides of the grid in model cwb4 corresponds to 
$3.2-6.86\,D_{\rm sep}$ as the stars progress in their orbit from
apastron to periastron.}
\label{tab:grid}
\begin{tabular}{lllll}
\hline
\hline
Model & No. of cells & \multicolumn{2}{c}{Length of sides} & Resolution \\
      &              & $(\Rsol)$ & $(a)$ & $(\Rsol)$ \\
\hline
cwb1 & $480^{3}$ & 240 & 7 & 0.5 \\
cwb2 & $456^{3}$ & 570 & 7.46 & 1.25 \\
cwb3 & $456^{3}$ & 570 & 7.46 & 1.25 \\
cwb4 & $480^{3}$ & 240 & 4.34 & 0.5 \\
\hline
\end{tabular}
\end{center}
\end{table}

A companion star influences the driving of a wind in a number of
distinct ways. Very close to the surface of the star, reflection
of the companion's starlight from the stellar photosphere can actually
enhance the wind driving and mass loss, and the angular asymmetry in
the line-scattering probability can induce significant tilting of the
direction vector of the wind away from the surface normal
\citep{Gayley:1999}.  Reflection effects become less important further
from the star. At this point the main effect of the companion's
radiation field is to reduce the net radiative flux, slowing the
remaining acceleration and producing a lower terminal velocity - this
effect is known as radiative inhibition \citep{Stevens:1994}. The
gravity of the companion star also has an effect - it alters the net
force on the wind, and can distort the shape of the other star, which
may lead to gravity darkening of the stellar surface and thus again
affect the wind.

Outflow boundary conditions are used for all of the simulations. The
fluid values on the grid are initialized by mapping 1D solutions of
the CAK equations onto the grid. The wind velocities are modified to
account for the velocity of each star on the grid (i.e. ${\bf v} =
v_{\rm w}({\bf r}) + v_{*}$). We assume that the stars remain closely
spherical and are not subject to significant deformations due to the
companion's gravity. To generate the winds we re-initialize density,
pressure and velocity values within a shell of 3-cells thickness
around each star at the start of every timestep. This procedure means
that the stars maintain a constant mass-loss rate, and have a
wind which is initially directed radially outwards.  Because our
models do not resolve the wind acceleration regions very close to the
stars, we are unable to investigate reflection effects on the
mass-loss and wind dynamics.  Instead the focus is on exploring the
latter stages of the wind acceleration and the effects this has on the
resulting wind-wind collision.

Our treatment of the wind initiation also means that the winds
effectively feel no radiative inhibition effects until outside the
3-cell remap radius. While this is not entirely satisfactory, our
attempts to self-consistently accelerate the winds from the stellar
surfaces have to date always resulted in unacceptable wind structure
due to a ``staircase'' effect introduced by the grid zones which
straddle the surfaces of the stars. This problem is very much reduced
in 2D $r-\theta$ simulations of single stars when the sonic point of
the winds is resolved \citep[see e.g.][]{Owocki:1994}. To resolve this
problem in 3D simulations with more than one star may require
adaptive-mesh-refinement (AMR) calculations, and/or sophisticated grid
geometries. For the time being we proceed with the above limitations,
with the aim of reducing/eliminating this problem in future work.

We specify the initial boundary between the winds as a flat 2D plane
which is normal to the line of centers through the stars and passes
through the stagnation point where momentum balance is achieved. This
position is calculated assuming the winds instantaneously reach
terminal velocity.  The simulations are evolved until the effects of
the initial conditions have been driven off the grid, following which
we start our analysis. For numerical reasons the orbital plane was aligned
with the $yz$ plane on the grid, with the $x$-axis perpendicular to this.

\subsection{Postshock equilibration}
\label{sec:postshock_eq}
There is now strong evidence from studies of supernova remnants that
the postshock thermalization of electrons lags behind that of ions in
high speed collisionless shocks, with the ratio of postshock electron
to proton temperature, $T_{\rm e}/T_{\rm p}$, substantially below unity
for shock velocities $\gtsimm 1000\kmps$
\citep{Rakowski:2005,vanAdelsberg:2008}.  Temperature equilibration
through Coulomb collisions and plasma instabilities then occurs some
distance downstream. One may expect these findings to be of relevance
to collisionless shocks in general and thus also to CWBs. In fact, the
postshock thermalization of electrons has already been examined in the
prototypical colliding winds binary WR\thinspace140 by
\citet{Zhekov:2000} and by \citet{Pollock:2005}, and can naturally
account for the softer-than-expected X-ray continuum emission
\citep[see also][]{Pittard:2006}.  

Despite an increased understanding of the need to consider slow
electron heating, the physics of the heating process in the shock
layer remains poorly understood. For instance, the appropriate value
of $T_{\rm e}/T_{\rm p}$ immediately downstream of the subshock is not
clear, with observations yielding a minimum value of $\sim 0.03$ at
shock velocities $\gtsimm 1500\kmps$ which is greater than the
theoretical minimum ($m_{\rm e}/m_{\rm p}$) by two orders of
magnitude, but smaller than the predictions of some collisionless
heating models \citep[e.g.][]{Cargill:1988}. Faced with this
situation, we have therefore assumed in our models that $T_{\rm
e}/T=0.2$ immediately postshock, where $T$ is the mean plasma
temperature \citep[see also][]{Zhekov:2000}.  The change in the
downstream electron temperature is then given by \citep{Spitzer:1978}
\begin{equation}
\label{eq:tet}
\frac{d(T_{\rm e}/T)}{dt} = 3.8 \times 10^{-12} (n_{\rm e} + n_{\rm i}) \left(1 - \frac{T_{\rm e}}{T}\right) \frac{{\rm ln}\;\Lambda}{30} \left(\frac{T_{\rm e}}{10^{7}\;{\rm K}}\right)^{-3/2}\;{\rm s}^{-1},
\end{equation}
where the numerical constant assumes cosmic abundances and ${\rm
ln}\;\Lambda \approx 30$ \citep*[cf.][]{Borkowski:1994}. Equipartition
requires that the product of the postshock timescale and the density
exceed $nt \gtsimm 3 \times 10^{11} \,{\rm
cm^{-3}\;s}$. Clumping in the winds will also
affect the electron thermalization timescale \citep{Pittard:2007}.
Eq.~\ref{eq:tet} is solved along with the other
hydrodynamical equations.

\begin{table*}
\begin{center}
\caption[]{Some key parameters of each model. In model cwb3 two values
are given for many of the parameters - the first corresponds to the
primary (O6V) star/wind, and the second to the secondary (O8V) star/wind. For
model cwb4 values at periastron and apastron are noted. $v_{\rm orb}$
is the orbital speed of the stars and $v_{\rm w}$ is the preshock wind
speed along the line-of-centres, both in $\kmps$. $v_{\rm w}/v_{\rm
s}$ is the ratio of the preshock wind speed along the line-of-centres
to the wind speed at the same radial distance from a single-star
solution.  Smaller values indicate more effective radiative
inhibition.  The ratio $v_{\rm orb}/v_{\rm w}$ affects the aberration
angle, $\theta_{\rm ab}$, of the WCR. The values quoted for
$\theta_{\rm ab}$ are measured directly from the simulations (not
calculated from the simple formula in Section~\ref{sec:models}), and
noted in degrees. The degree of downstream curvature of the WCR in the
orbital plane is specified by $\alpha_{\rm cor}$, where the
curvature is assumed to trace an Archimedean spiral which in polar
coordinates is described by $r = \alpha_{\rm cor}\theta$. The
value of $\alpha_{\rm cor}$ corresponds to the downstream
distance (in units of $d_{\rm sep}$) along the WCR for each radian of
arc it sweeps out in the orbital plane. Smaller values indicate
tighter curvature.  The leading and trailing arms of the WCR in model
cwb3 display differing degrees of curvature, so the value quoted for
this model is a rough average. For model cwb4 the quoted values
roughly represent the ``instantaneous'' curvature at periastron and
apastron. $\chi$ is the ratio of the postshock cooling to flow
timescales, and is calculated using the pre-shock wind speed, $v_{\rm
w}$, as measured from the simulations.  $\chi \ltsimm 1$ indicates
that the shocked gas rapidly cools, while $\chi \gtsimm 1$ indicates
that the plasma in the WCR remains hot as it flows out of the system.}
\label{tab:keyparams}
\begin{tabular}{llllllll}
\hline
\hline
Model & $v_{\rm orb}$ & $v_{\rm w}$ & $v_{\rm w}/v_{\rm s}$ & $v_{\rm orb}/v_{\rm w}$ & $\theta_{\rm ab}$ & $\alpha_{\rm cor}$ & $\chi$ \\
\hline
cwb1 & 290 & 730 & 0.58 & 0.40 & 17 & 3.5 & 0.34 \\
cwb2 & 225 & 1630 & 0.90 & 0.14 & 3-4 & 6.5 & 19 \\
cwb3 & 152,208 & 1800,1270 & 0.90,0.88 & 0.08,0.16 & 2 & 4.5 & 28,14 \\
cwb4 & 334-156 & 710-1665 & 0.57-0.91 & 0.47-0.09 & 21-4 & 3-10 & 0.34-19 \\
\hline
\end{tabular}
\end{center}
\end{table*}

The timescale for the postshock ionization to approach equilibrium
will also be an important factor in some CWBs. For instance,
non-equilibrium ionization may explain why line-profile models which
assume rapid ionization equilibrium \citep{Henley:2003} are unable to
reproduce the observed correlation of X-ray line widths with
ionization potential in the Wolf-Rayet system $\gamma$-Velorum
\citep{Henley:2005}. Although the postshock ionization depends on the
thermal history of the plasma, to zeroeth order the ionization is
independent of the temperature history and the specific ionic species
in the plasma \citep{Masai:1994}, and can be characterized by the
so-called ionization age, $n_{\rm e}t$. For example, the ionization of
oxygen and iron in plasma at $T=10^{6.5}\,$K as a function of $n_{\rm
e}t$ is shown in Figs.~1 and~2 of \citet{Hughes:1985}. The oxygen
reaches ionization equilibrium at $n_{\rm e}t \approx 10^{12.5}\;{\rm
cm^{-3}\;s}$, while the iron reaches equilibrium at $n_{\rm e}t
\approx 10^{11.5}\;{\rm cm^{-3}\;s}$.  At higher temperatures, iron is
ionized to higher stages, and requires a greater ionization age to
reach equilibrium. However, in general a plasma will be close to
ionization equilibrium when $n_{\rm e}t \gtsimm 10^{12.5}\;{\rm
cm^{-3}\;s}$.  We therefore advect a scalar variable in the
hydrodynamical code to track the postshock value of $n_{\rm e}t$.

\subsection{Models investigated}
\label{sec:models}
In this work we do not attempt to model particular systems. Instead
the aim is to achieve an understanding of how the dynamics of the
collision region depend on some key parameters. We have therefore
computed 4 different models with a range of binary and stellar
parameters, as summarized in Tables~\ref{tab:models}
and~\ref{tab:stellar_params}.  In the present work we consider systems
with dwarf O-stars, so that tidal distortions are minimized, and in
this first study can be ignored.  The stellar parameters are
consistent with recent observations presented in
\citet{Repolust:2004}, \cite{Martins:2005}, \citet{Bouret:2005} and
\citet{Fullerton:2006}, in which clumping within the winds is
considered. The CAK parameters ($\alpha$ and $k$) which are needed to
drive the desired winds from the stars are noted in
Table~\ref{tab:stellar_params}, along with the value of $\bar{Q}$ 
calculated from them. We assume that the mass-loss from the
stars is isotropic. Solar abundances are assumed for each star/wind,
and all simulations are performed in the centre of mass frame.

The nature of the WCR is largely governed by the ratio of the cooling
timescale for the shocked gas to the dynamical timescale for it to
flow out of the system, which is given by $\chi =
v_{8}^{4}d_{12}/\Mdot_{-7}$, where $v_{8}$ is the preshock velocity in
units of $1000\;\kmps$, $d_{12}$ is the stellar separation in units of
$10^{12}\;{\rm cm}$, and $\Mdot_{-7}$ is the stellar mass-loss rate in
units of $10^{-7}\;\Msolpyr$ \citep{Stevens:1992}.  Another key
parameter governing the nature of the WCR is the ratio of the
orbital speed to the pre-shock wind speed, $v_{\rm orb}/v_{\rm
w}$. This ratio affects the aberration angle of the WCR [$\theta \sim
\tan^{-1} (v_{\rm orb}/v_{\rm w})$] and its degree of downstream
curvature due to coriolis forces. 

In this work we consider systems with identical or similar wind
momenta. This means that radiative braking \citep{Gayley:1997} effects
are negligble, and in all cases a normal wind-wind pressure balance
exists. In future work we will extend our investigation to systems
with higher wind momentum ratios, to study the effects of braking and
the dynamics of the WCR when the more powerful wind overwhelms the
weaker wind and the WCR collapses onto the surface of one of the
stars.

In model cwb1 we suppose that two identical O6V stars move around each
other in a circular orbit with a period of 3 days. The stellar
separation is $34.26\;\Rsol$, and each star has an orbital velocity
$v_{\rm orb} = 290\;\kmps$.  A rough estimate of $\chi$ can be
obtained by determining the speed of the winds at the stagnation point
where momentum balance is achieved. Ignoring for the moment effects
due to the orbital motion of the stars and the presence of two
radiation fields, one finds that the winds accelerate to a speed of
$1250\;\kmps$ prior to their collision at the stagnation point.  This
implies that $\chi=2.9$, and the postshock gas would not be expected
to cool back to $T\sim10^{4}\;$K until it had left the central regions
of the system. In fact, we shall see that once the simulation starts,
the reduction in the net radiative flux due to the presence of the
companion star reduces the acceleration of each wind towards its
companion. This ``radiative inhibition'' \citep{Stevens:1994} reduces
the pre-shock velocities below $1250\;\kmps$, which increases the
cooling in this model (see Table~\ref{tab:keyparams}, which summarizes
some of the salient features directly obtained from the hydrodynamcial
simulations).  Neglecting radiative inhibition again, we also find
that $v_{\rm orb}/v_{\rm w} = 0.23$, indicating that the WCR will
display a reasonably large aberration angle and downstream curvature
due to coriolis forces. These effects become more pronounced if there
is a significant reduction in wind speeds along the line-of-centres
due to raditive inhibition. Note also that unphysical masses are
required in some of the models of \citet{Lemaster:2007}, including
their main large-box simulation C2.5.

\begin{figure*}
\psfig{figure=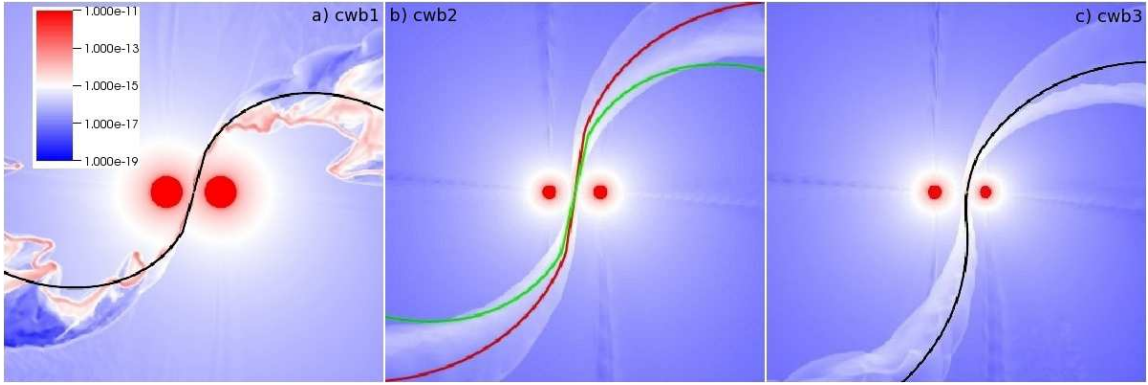,width=15.2cm}
\caption[]{Density plots of models cwb1 (a), cwb2 (b), and cwb3
(c). The colour scale is logarithmic, spanning $10^{-19}\gpcm3$ (blue)
to $10^{-11}\gpcm3$ (red). Panel (a) has sides of length $240\;\Rsol$,
while panels (b) and (c) have sides of length $570\;\Rsol$. The black,
red and green lines marked on the figures show the position of the
contact discontinuity as calculated by the model described in
\citet{Parkin:2008} - see the text for further details.}
\label{fig:cwb123_density}
\end{figure*}

\begin{figure*}
\psfig{figure=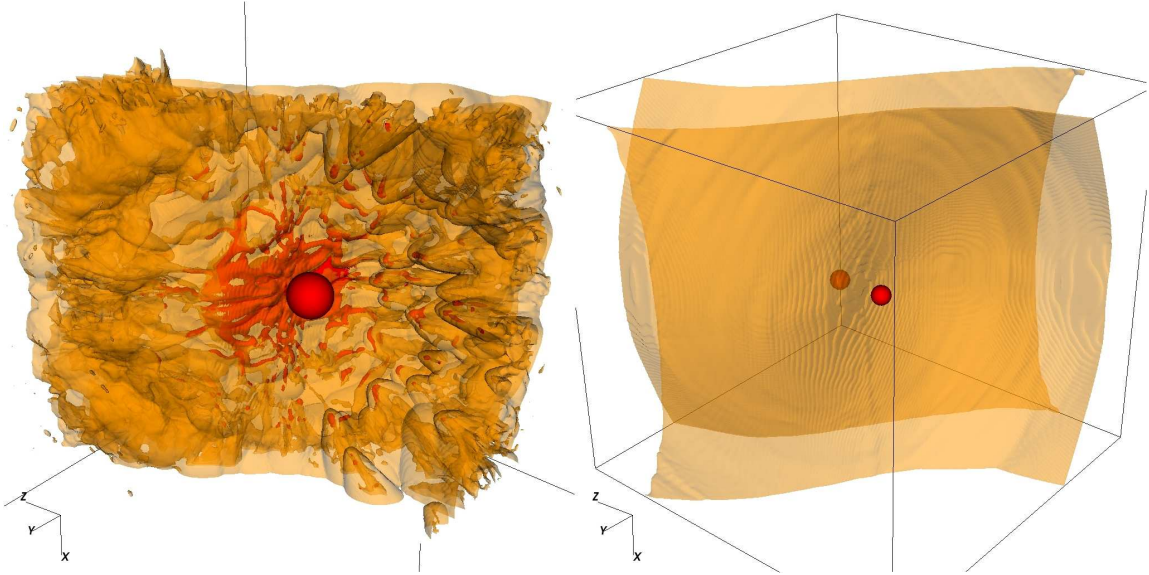,width=15.2cm}
\caption[]{a) 3D density contours from model cwb1. The orange contour
shows the surface at a temperature of $3 \times 10^{5}$\,K, while the
red contour shows the surface at a density of $10^{-14}\;{\rm g \,
cm^{-3}}$. The density contour highlights the stellar positions, plus
the densest regions of the WCR (the other star is hidden behind
this). The WCR readily breaks up into small dense clumps on the
trailing edge of each arm. These clumps are often surrounded by high
temperature bow shocks as revealed by the temperature surface.  b) As
Fig.~\ref{fig:cwb12_3ddensity2}(a) but for model cwb2. The shocks
trace a double-helix, in a similar manner to the DNA molecule.}
\label{fig:cwb12_3ddensity2}
\end{figure*}

\begin{figure*}
\psfig{figure=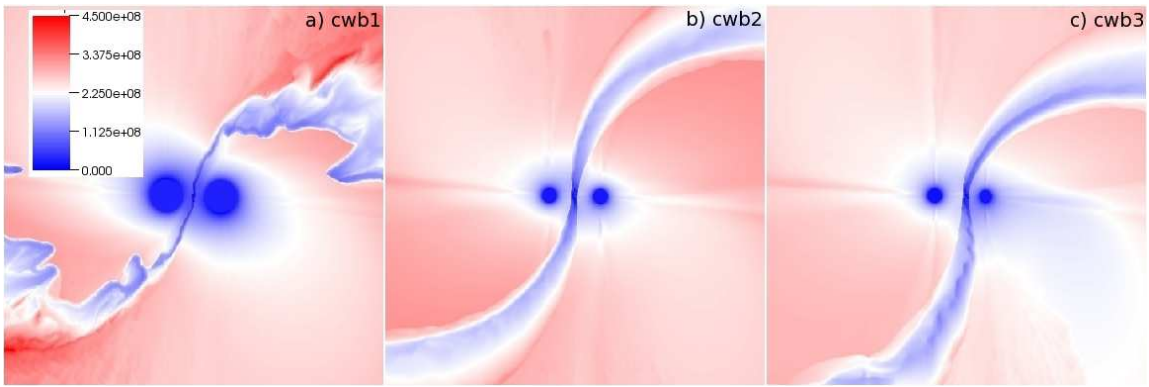,width=15.2cm}
\caption[]{Velocity plots of models cwb1 (a), cwb2 (b), and 
cwb3 (c) in the orbital plane. The colour scale is linear, spanning
$0\;\kmps$ (blue) to $4500\;\kmps$ (red).}
\label{fig:cwb123_velocity}
\end{figure*}

\begin{figure*}
\psfig{figure=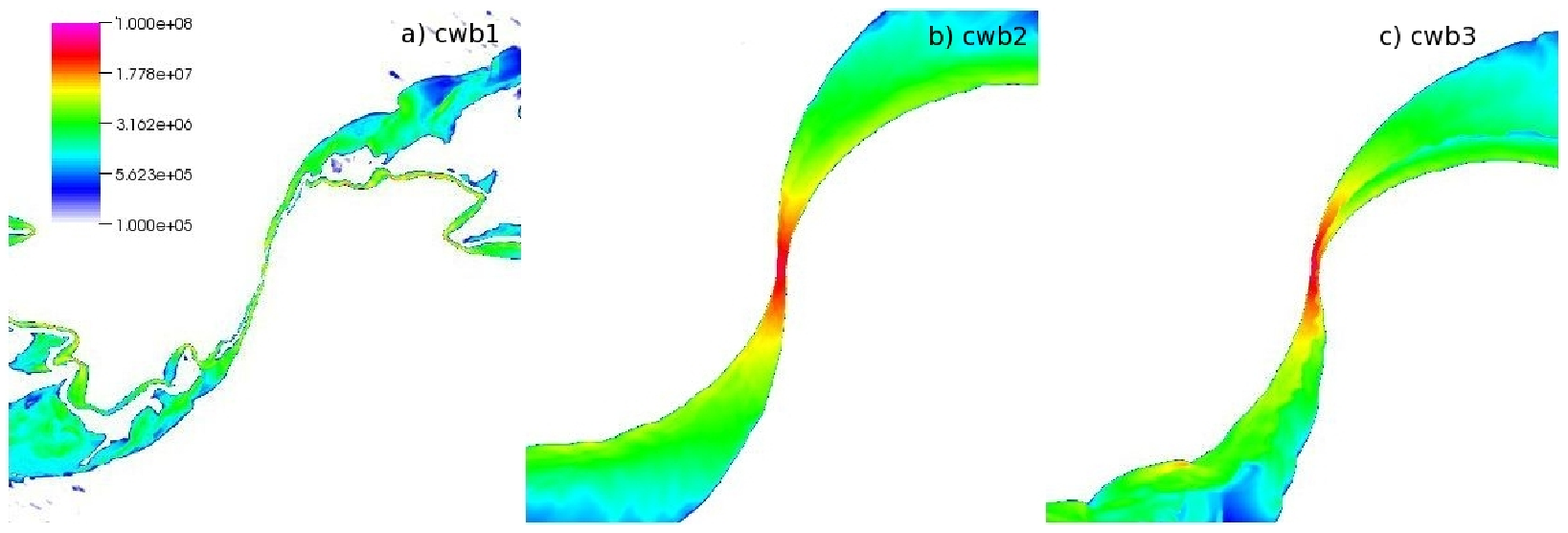,width=15.2cm}
\caption[]{Temperature plots of models cwb1 (a), cwb2 (b), and 
cwb3 (c) in the orbital plane. The colour scale is logarithmic, spanning
$\leq 10^{5}\;$K (white) to $10^{8}\;$K (pink).}
\label{fig:cwb123_temp}
\end{figure*}

Model cwb1 is similar to many real systems, including DH\,Cep
\citep[see][and references therein]{Linder:2007}, HD\,165052
\citep{Arias:2002,Linder:2007}, and HD\,159176
\citep{DeBecker:2004b,Linder:2007}. All of these systems have near
identical main-sequence stars of spectral type O6-O7, and circular or
near-circular orbits with periods near 3 days. Hence the hydrodynamics
of and emission from the WCR in model cwb1 will be a reasonable
approximation to the situations in these systems.  In some of these
systems it may be difficult for the winds to attain a normal ram-ram
pressure balance, and the stronger wind may push the WCR onto the
surface of the star with the weaker wind.  Detailed simulations of
each individual system are needed to ascertain whether, in fact, this
occurs.  Reflection and distortion effects may also well be important
in these systems, but are beyond the scope of the present work.

Model cwb2 uses the same stars as model cwb1, but increases the
orbital period to 10 days (the stellar separation becomes $76.3\;\Rsol$).
The pre-shock collision speed on the line of centers (neglecting
radiative inhibition) increases to $1970\;\kmps$, as the winds now 
have more room to accelerate. The higher collision speed leads to
higher postshock temperatures, and together with the increased 
stellar separation produces lower postshock densities. The result is
that cooling within the WCR is dramatically
reduced ($\chi \approx 40$), and the postshock gas behaves largely 
adiabatically. The reduced orbital speed and higher pre-shock
wind speeds also means that the aberration and curvature of the WCR
will be smaller in this model than in model cwb1. Taking account
of radiative inhibition we again find slightly increased cooling and
orbital effects (see Table~\ref{tab:keyparams}).

Model cwb2 is similar to HD\,93161A, an O8V + O9V system with a
circular orbit and an orbital period of 8.566 days \citep{Naze:2005},
albeit with slightly more massive stars and powerful winds. Another
system with not too dissimilar properties is Plaskett's star
\citep{Linder:2006,Linder:2008}, though this object contains stars
which have evolved off the main sequence. The wider separation of the
stars in these systems means that the dominant dynamical effect of the
companion starlight will be radiative inhibition, with reflection
effects minimized.

In model cwb3, we examine the interaction of unequal winds in a
hypothetical O6V + O8V binary. We keep the same stellar separation as
model cwb2, which gives a period of $10.74\;$days from the
total system mass of $52\Msol$. The orbit is circular, with the
O8V star further from the center of mass. An initial estimate of the
preshock wind speeds can be obtained by considering the location
of the stagnation point. The terminal speed wind
momentum ratio, $\eta = \Mdot_{2}v_{\infty,2}/\Mdot_{1}v_{\infty,1} =
0.4$, indicating that the stagnation point occurs
at a distance of $0.39\;a$ from the O8V star. 
At this distance the secondary wind speed is $1570 \kmps$, giving
$\chi_{2} \approx 32$.  The primary wind collides at higher speed
($2070 \kmps$), and its postshock plasma is slightly more adiabatic
($\chi_{1} \approx 50$).  Again, radiative
inhibition reduces the preshock wind speeds and postshock cooling 
parameters below these simple estimates (see Table~\ref{tab:keyparams}).

Model cwb4 explores the effect of an eccentric orbit which takes the
stars through separations of $34.26-76.3\;\Rsol$ (i.e.  the
separations of the stars in the circular orbits of models cwb1 and
cwb2).  The required eccentricity is $e=0.\overline{36}$, and the
resulting orbital period is 6.1~days.  This model allows us to
investigate whether the WCR properties at a specific stellar
separation are similar to those obtained with a circular orbit.  Some
well-known O+O binaries with eccentric orbits include (in order of
increasing orbital period) HD\,152248 \citep[$e =
0.127$;][]{Sana:2004}, HD\,93205 \citep[$e=0.46$;][]{Morrell:2001},
HD\,93403 \citep[$e=0.234$;][]{Rauw:2002}, Cyg\,OB2\#8A
\citep[$e=0.24$;][]{DeBecker:2004,DeBecker:2006} and $\iota$\,Orionis
\citep[$e=0.764$;][]{Bagnuolo:2001}.

The hydrodynamical grid is half-cubic in all simulations, and is
reflected in the orbital plane for further analysis.  The grid is
large enough to capture, for example, the majority of the X-ray
emission from each model.  Details of the grids are noted in
Table~\ref{tab:grid}.  As a comparison, most of the simulations run by
\citet{Lemaster:2007} used a grid with sides of length $2.5\;a$, with
their large grid simulations having sides of length $6.25\;a$.

\section{Results}
\label{sec:results}

\subsection{Model cwb1}
\label{sec:dynamics_cwb1}
Fig.~\ref{fig:cwb123_density}(a) shows a density snapshot in the
orbital plane of model cwb1. The postshock gas in the WCR between the
stars has cooled into a thin, dense sheet, which is subject to the
non-linear thin-shell instability
\citep{Vishniac:1994,Blondin:1995}. Clearly our initial estimate of
$\chi$ in Section~\ref{sec:models} was too high. Examination of the
dynamics reveals that the winds in fact collide at $\approx 730
\kmps$, rather than the $1250 \kmps$ obtained from considering the
acceleration of the wind of an isolated star. This is caused by a
significant reduction in the acceleration of the stellar winds towards
the centre of mass of the system due to strong radiative inhibition
(in reality reflection effects could also be important, though this
process is not examined in this work).
The cooling parameter, $\chi$, is reduced to $\approx 0.3$, which
gives rise to the severe radiative cooling of the postshock gas seen
in Fig.~\ref{fig:cwb123_density}(a).

As previously noted, the nature of the WCR is also determined by the
ratio of the pre-shock wind speed to the orbital speed of the stars,
since this affects its angle of aberration and downstream
curvature. The aberration angle of the WCR in
Fig.~\ref{fig:cwb123_density}(a) is $\approx 17^{\circ}$, which
compares to an expected value of $22^{\circ}$ if $v_{w} = 730 \kmps$
is used in the formula in Section~\ref{sec:models}.  That the expected
value is overestimated is entirely anticipated, of course, since the
pre-shock wind speed is higher off-axis than it is along the
line-of-centres, so that using speeds along the line-of-centres biases
the calculation to larger angles. 

Downstream the WCR spirals around the stars as a result of their
orbital motion. Analyses of the so-called ``pinwheel nebulae'' have
compared such morphologies to an Archimedean spiral, which in polar
coordinates is specified by $r=\alpha_{\rm cor}\theta$ \citep[see][and
references therein]{Tuthill:2008}. The curvature of the spiral is
governed by the distance which the plasma in the WCR moves during one
orbital period. The gas in the denser parts of the WCR near the right
edge of the grid shown in Fig.~\ref{fig:cwb123_density}(a) has a speed
of about $2050\;\kmps$. Hence in 3 days this travels a distance of
$3.55\;{\rm au}$ ($763\;\Rsol$). The minimum distance from the centre of the
grid to its edge is $120\,\Rsol$, so we expect the WCR to curve
through 16 per cent of a full rotation, or about $57^{\circ}$.
Fig.~\ref{fig:cwb123_density}(a) actually displays greater curvature
than this estimate, though this is expected because the speed of gas
near the apex of the WCR is much slower than further downstream.

\begin{figure*}
\psfig{figure=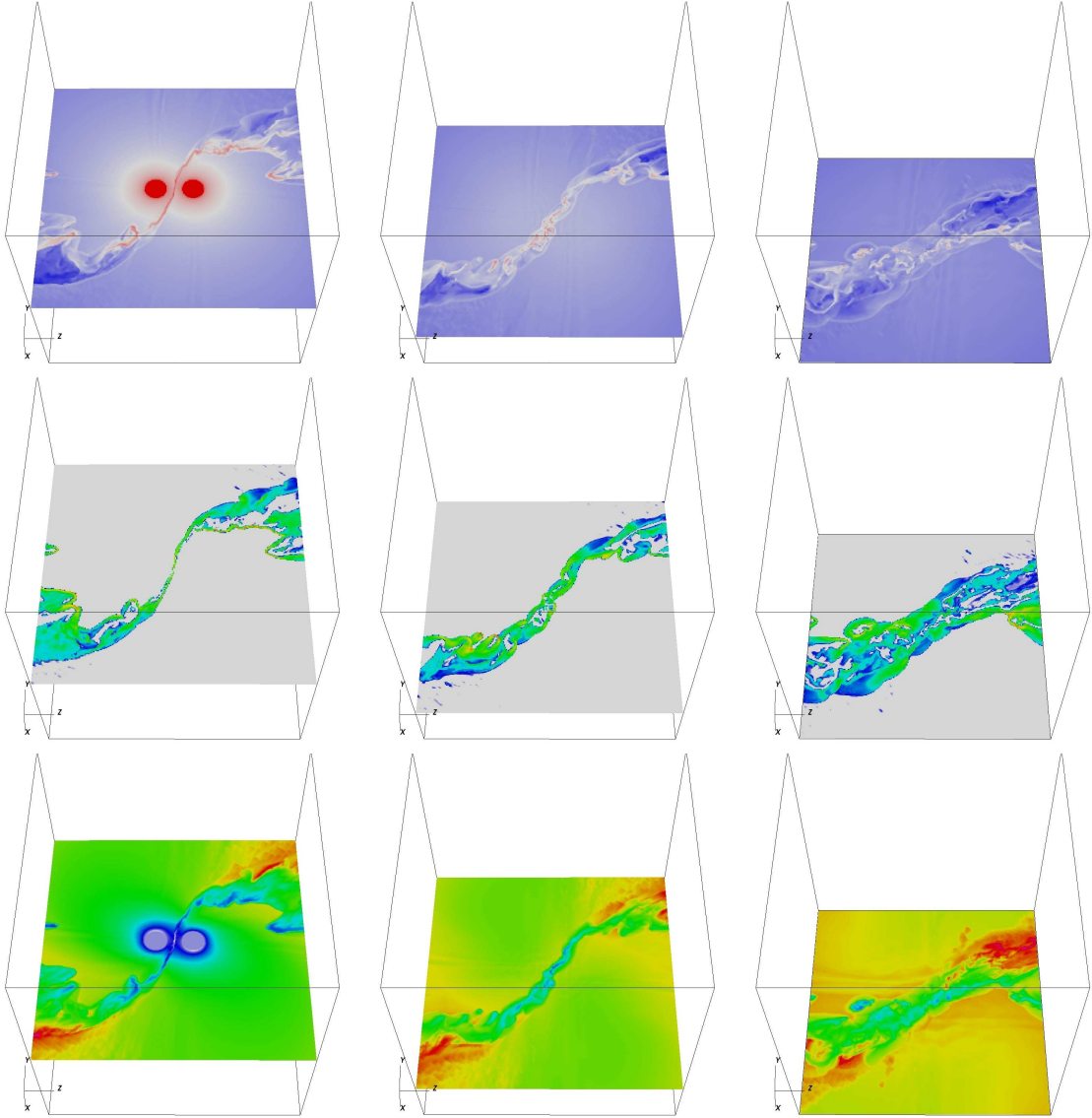,width=15.2cm}
\caption[]{Density (top), temperature (middle), and speed (bottom)
slices of model cwb1 in the orbital plane (left), and below the
orbital plane (middle and right). The density and temperature colour
scales are the same as in Figs.~\ref{fig:cwb123_density}
and~\ref{fig:cwb123_temp}, while the colour scale for the speed plots
is the same as the temperature scale though is linear and covers the
range $0-4500\kmps$ (i.e.  stationary/low speed gas is white/blue,
while the high speed gas is red).}
\label{fig:cwb1_3d_sliceyz_montage}
\end{figure*}

\begin{figure*}
\psfig{figure=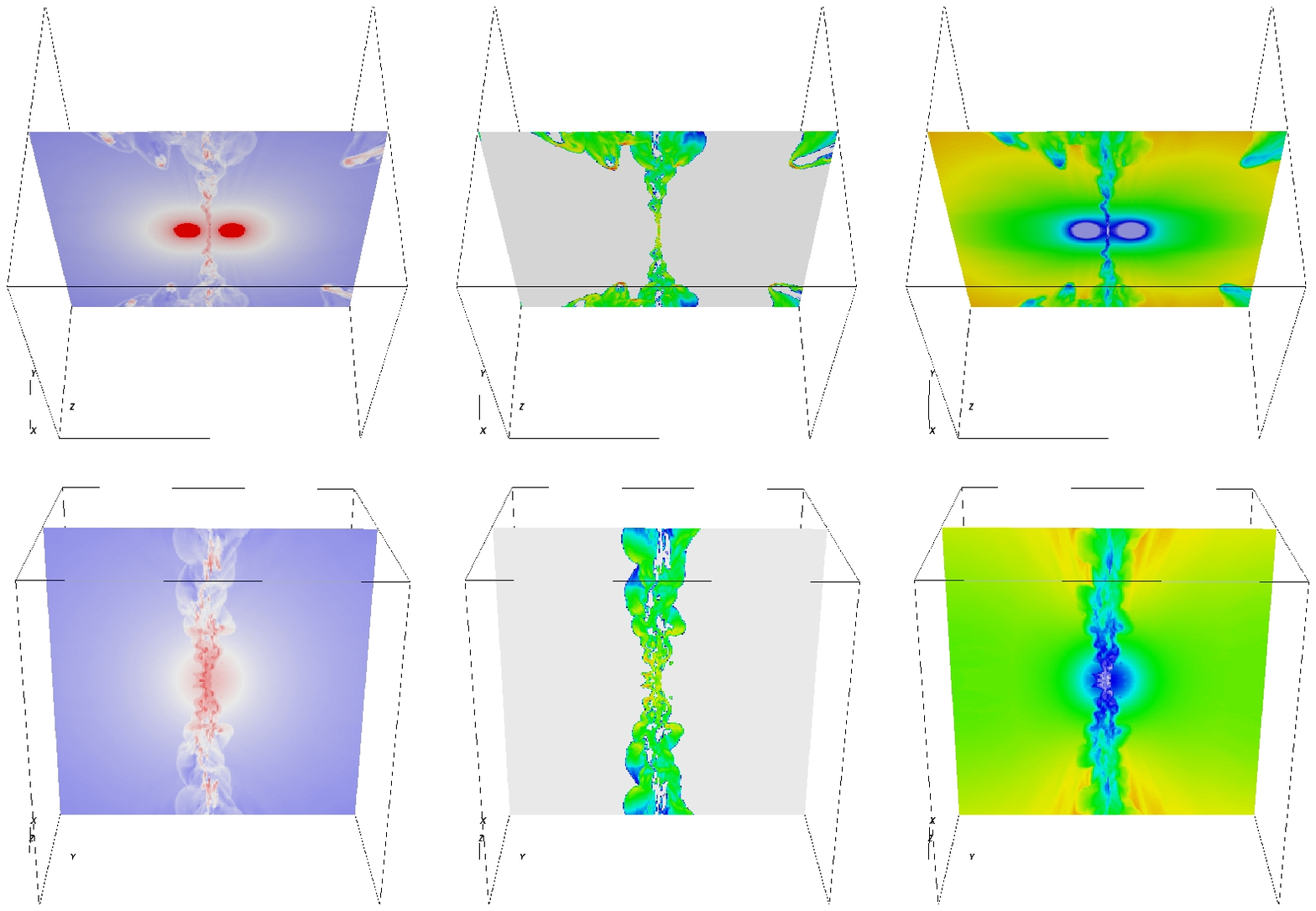,width=15.2cm}
\caption[]{Density (left), temperature (middle), and speed (right) 
slices of model cwb1 in the plane orthogonal to the orbital
plane containing the stars (top), and in the plane orthogonal to this
and the orbital plane (bottom). The colour scales are the same as in
Fig.~\ref{fig:cwb1_3d_sliceyz_montage}.}
\label{fig:cwb1_3d_slicexzxy_montage}
\end{figure*}

\begin{figure*}
\psfig{figure=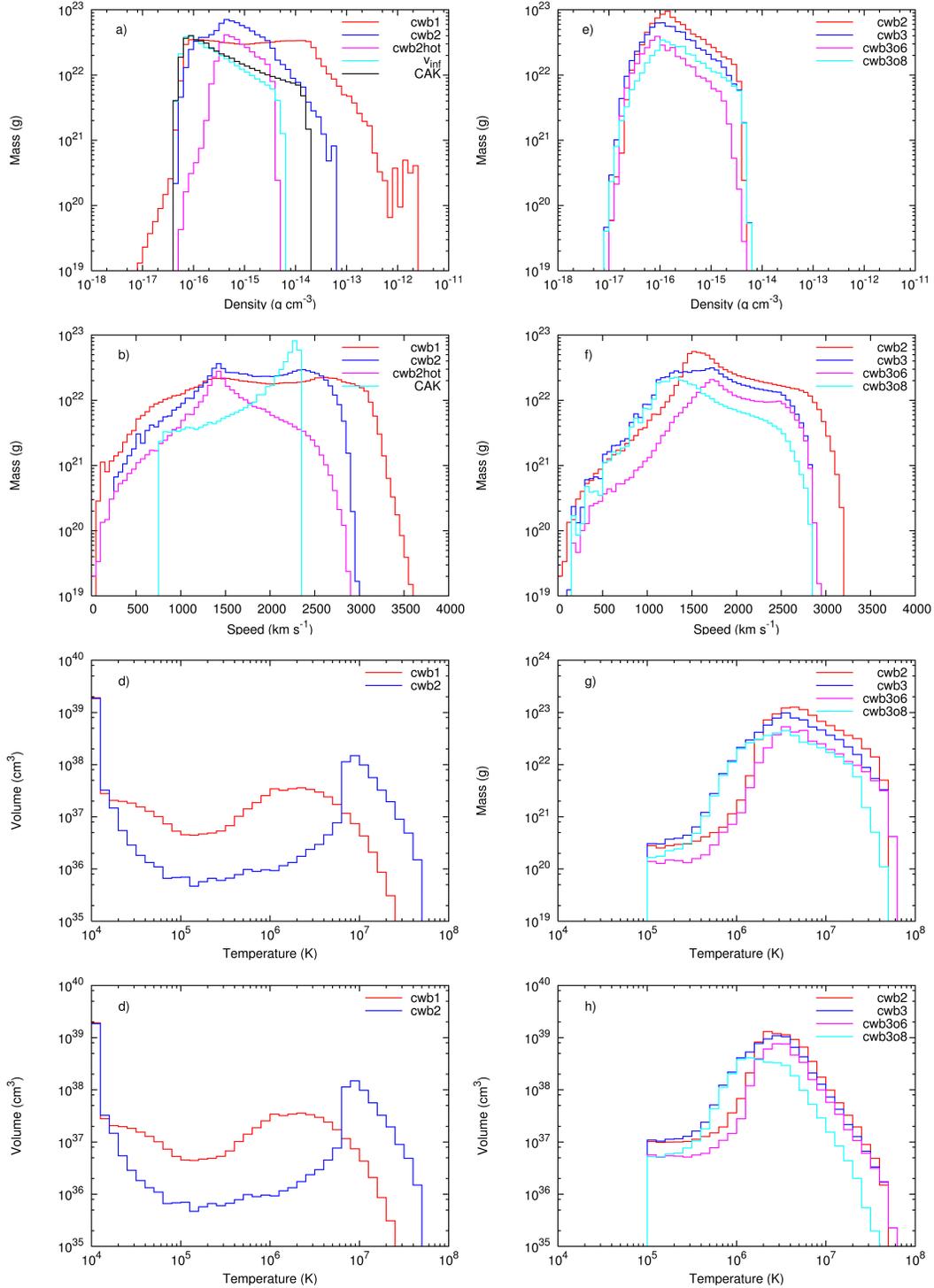,width=14cm}
\caption[]{Histograms of various quantities in the models. The left
column shows data from models cwb1 and cwb2 within a radius of
$120\;\Rsol$ of the system centre of mass, while the right column
shows data from models cwb2 and cwb3 within $285\,\Rsol$ of the centre
of mass. Plot a) shows mass versus density, plot b) mass versus speed,
plot c) mass versus temperature, and plot d) volume versus
temperature. Plot e)-h) show the equivalent analysis of models cwb2
and cwb5.}
\label{fig:statistics}
\end{figure*}

\begin{figure*}
\psfig{figure=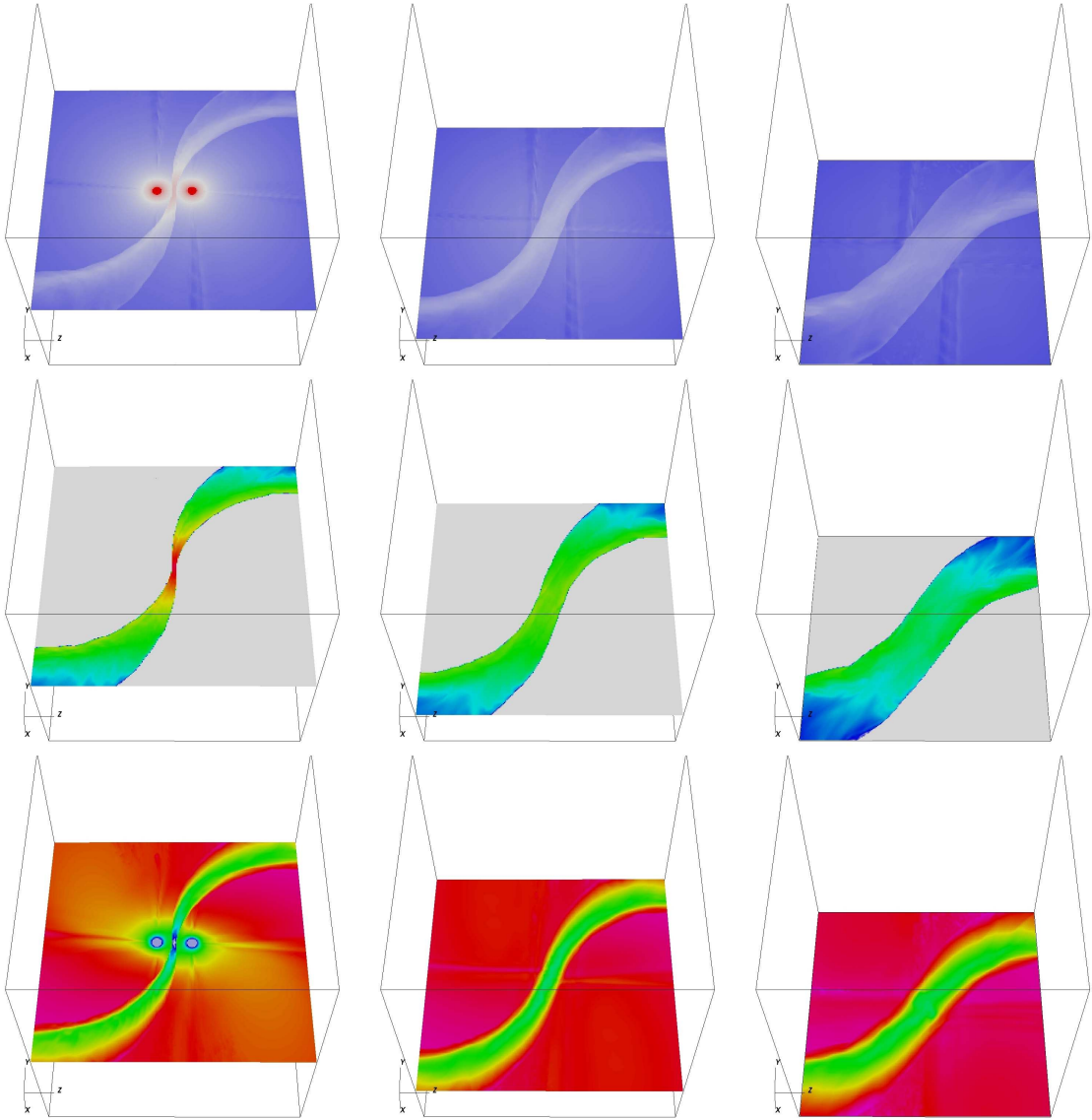,width=15.2cm}
\caption[]{Density (top), temperature (middle), and speed (bottom) 
slices of model cwb2 in the orbital plane (left), and below 
the orbital plane (middle and right). The colour scales are the same as in
Fig.~\ref{fig:cwb1_3d_sliceyz_montage}.}
\label{fig:cwb2_3d_sliceyz_montage}
\end{figure*}

\begin{figure*}
\psfig{figure=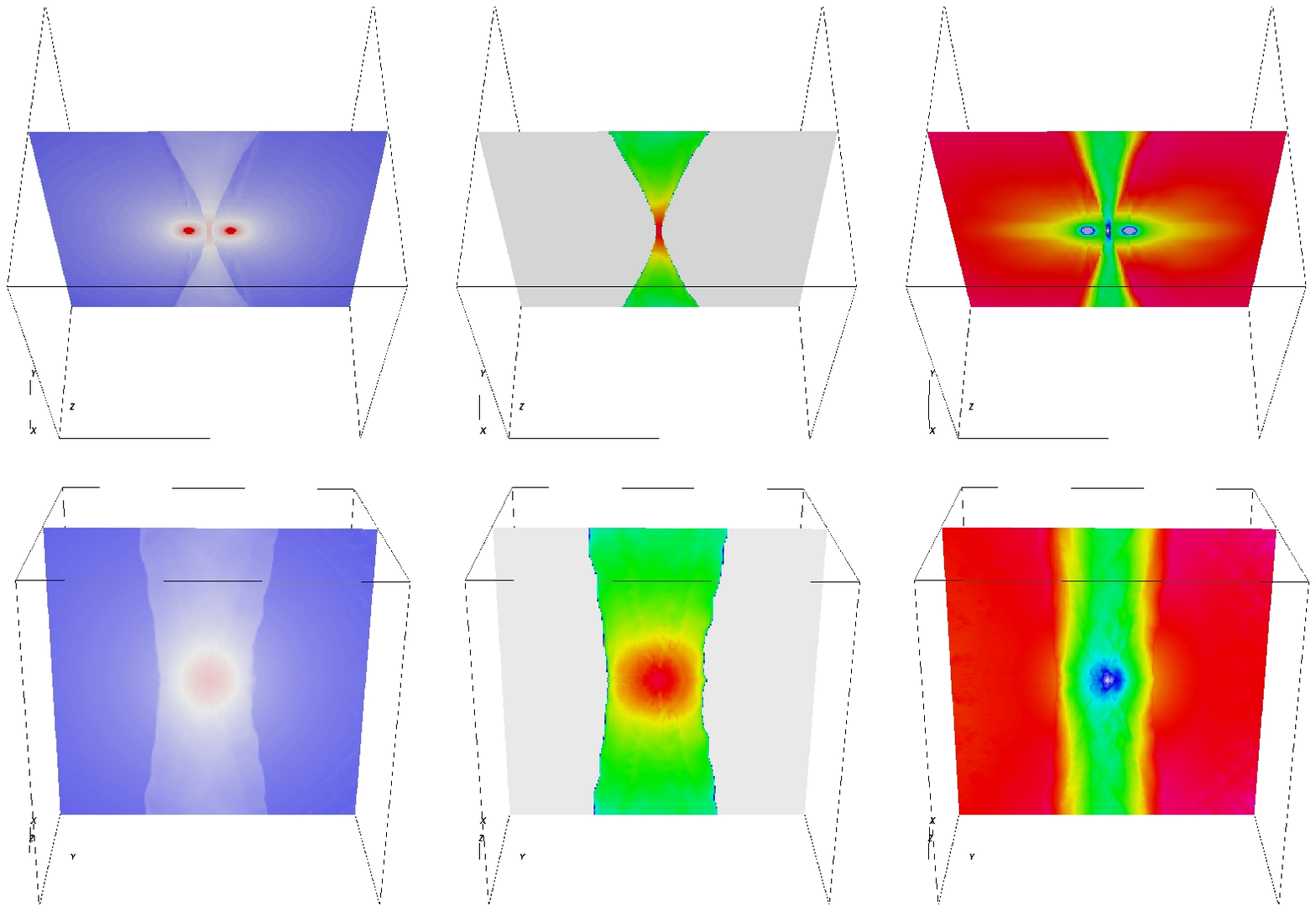,width=15.2cm}
\caption[]{Density (left), temperature (middle), and speed (right) 
slices of model cwb2 in the plane orthogonal to the orbital
plane containing the stars (top), and in the plane orthogonal to this
and the orbital plane (bottom). The colour scales are the same as in
Fig.~\ref{fig:cwb1_3d_sliceyz_montage}.}
\label{fig:cwb2_3d_slicexzxy_montage}
\end{figure*}

Overplotted on Fig.~\ref{fig:cwb123_density}(a) is the contact
discontinuity in the orbital plane as calculated by the dynamical
model described in \citet{Parkin:2008}. The model assumes that the
winds collide at terminal velocity, which is obviously not the case
here.  Nevertheless, the model is able to match remarkably well the
position of dense gas in the WCR, if the wind speeds are assumed to be
$2000\;\kmps$ and the flow in the WCR behaves ballistically once
accelerated to 75 per cent of its terminal speed \citep[for further
details see][]{Parkin:2008}. Adopting a $2500\;\kmps$ terminal wind
speed in the dynamical model results in slightly less aberration and
downstream curvature, while specifying a $750\;\kmps$ terminal wind
speed results in much greater aberration and a much tighter downstream
curvature compared to the hydrodynamical models. Clearly, there is
scope for the dynamical model to be ``calibrated'' against full
hydrodynamical models. A major advantage of the dynamical model
over full hydrodynamical calculations is its low computational cost.

From movies of the simulation it is clear that the radiative
overstability occurs in model cwb1. High Mach number shocks are
susceptible to a global oscillation if the slope, $\alpha$, of the
cooling function (where the cooling function, $\Lambda(T)$, is
approximated as $\Lambda_{0} T^{\alpha}$), is less than a
critical value \citep[i.e. $\alpha \ltsimm 0.4$ - see
e.g.][]{Strickland:1995,Pittard:2005}. In this model the postshock gas
temperature $T \approx 7 \times 10^{6}\;$K corresponds to a
local slope $\alpha \approx 0$ on the cooling function. This is less
than the critical value $\alpha_{\rm cr} = 0.4$, below which the
overstability occurs.  As the gas cools below $\approx 3 \times
10^{6}\;$K, very low values of $\alpha$ are encountered
\citep[see][]{Pittard:2005}.  Mass-loading (from clump ablation and
evaporation) can moderate or even prevent thermal instabilities
\citep*{Pittard:2003b}, but clearly is not a significant effect here.

Further downstream, the leading edge of the WCR becomes shadowed by
upstream regions of the WCR, leading to very low post-shock
densities, and very weak oblique shocks. In contrast, the trailing
edge of the WCR is very much denser, and remains much more normal to
the wind. The WCR remains denser on the trailing side because of the
high inertia of the dense gas, which causes it to respond slowly to
the movement of the stars in their orbits. The dense sheet of cooled
material between the stars is broken up into numerous small clumps in
the downstream flow on the trailing side of the WCR, as shown in
Fig.~\ref{fig:cwb12_3ddensity2}(a) which shows 3D isosurfaces of the
density and temperature in model cwb1. In contrast, the trailing edge
of the WCR is far smoother and less structured.

Fig.~\ref{fig:cwb123_velocity}(a) shows the wind speeds in the orbital
plane. The driving of the winds off the back of the stars produces
wind speeds which are lower in the centre of mass frame than off the
front of the stars. Inspection of the model also reveals that the
winds accelerate more slowly when in the shadow of one of the stars.
Remarkably, the maximum speed of material on the computational grid in
the orbital plane is $\approx 4650 \kmps$, almost double the terminal
velocity of the single star wind. However, these speeds occur in
extremely rarefied gas within the WCR and are atypical of the highest
speeds seen. It is unlikely that such high speeds would be easily
detected since they are tied to such small amounts of material. A more
representative speed of the high velocity gas is $3000-3500 \kmps$
(see Fig.~\ref{fig:statistics}b) - such speeds are typically seen in
the shadow of the leading shocks of the WCR. The direction of the
lowest wind speeds is towards the top left and bottom right of the
density plot in Fig.~\ref{fig:cwb123_density}(a), where the typical
wind speed is $\approx 2750 \kmps$. The winds are slower in these
directions because of the oppositely directed stellar motions. The
winds are still accelerating in these regions, albeit gradually.

Fig.~\ref{fig:cwb123_temp}(a) shows the temperature in the orbital
plane of model cwb1.  The expected postshock temperature behind a high
Mach number shock which is normal to the preshock flow is given by $T
= \frac{3}{16} \frac{\overline{m}}{k} v^{2}\;$K, where $\overline{m}$
is the average mass per particle. Given the observed preshock wind
speeds along the line of centres we expect a postshock
temperature $T = 7.2 \times 10^{6}\;$K, which is indeed what is seen.

In wide CWBs where the winds collide at their terminal speeds the
maximum postshock temperature is produced at the apex of the WCR where
the shocks are normal to the flow.  For a given preshock speed,
oblique shocks are less efficient at thermalizing the flow and produce
lower postshock temperatures.  However, in systems where the winds
collide prior to reaching their terminal speeds the hottest
temperatures in the WCR can in fact be obtained far downstream of the
WCR apex, as we now explain. The increased distance from the stars to
the shocks means that the winds have more time to accelerate and
collide at higher speed. While these increased speeds may be offset by
the increasing obliquity of the shocks, in cases where the WCR is
strongly radiative and thus susceptible to kinking by thin shell
instabilities, some parts of the WCR surface (and thus the shocks) can
be normal to the preshock flow.  This combination of high preshock
speeds and normal shocks can result in the hottest plasma temperatures
actually occuring far downstream. This is in fact the case in model
cwb1, where temperatures reach as high as $4.5\times10^{7}\;$K at the apex of
bowshocks around downstream clumps, indicating a relative speed of
$\approx 1800\;\kmps$ between the dense clumps and the preshock winds
at such points (though the typical temperature in these bowshocks is
somewhat lower - see Fig.~\ref{fig:statistics}b). 
Note that the maximum temperature is actually higher
than the corresponding value from model cwb2 (see
Section~\ref{sec:dynamics_cwb2}), despite the latter models wider
stellar separation and higher preshock wind speeds along the
line-of-centres (at first glance Fig.~\ref{fig:statistics}(b) would seem
to contradict this statement, but this is only because the ordinate as
plotted does not extend to low enough values).

In model cwb1, the densities between the stars are sufficiently high
for the shocks to be collisional. Hence, the electron and ion
temperatures rapidly come into equilibrium. The post-shock density
(prior to cooling) along the line-of-centres is $n \sim
10^{10}\;\pcm3$ (this increases to a maximum of $2.4 \times
10^{12}\;\pcm3$ after cooling - see Fig.~\ref{fig:statistics}a). Hence
the electron and ion temperatures reach equilibrium about 30 seconds
after passing through the shocks.  The equilibration timescale is
longer for gas shocked downstream of the WCR apex, but still very
rapid.

Similarly, the ionization age of the shocked plasma, $n_{\rm e} t$, exceeds
$10^{13} \;{\rm cm^{-3}\;s}$ near the apex of the WCR, indicating that
the post-shock gas is in collisional ionization equilibrium. Lower
ionization ages are seen further downstream (e.g. in the bowshocks
surrounding the protrusions of cold postshock gas seen downstream in
the trailing edge of the WCR in Fig.~\ref{fig:cwb12_3ddensity2}(a),
$n_{\rm e}t \sim 10^{11.5-12} \;{\rm cm^{-3}\;s}$). While the ionization will
not be in equilibrium in such regions, their densities are
sufficiently low that their emission will be a very minor contribution
to the total from the system, so, for instance, it is likely that
there will be no discernible effect on the global X-ray emission
(though, as already noted, these regions are also the hottest parts of
the WCR and thus produce the hardest emission).

Figs.~\ref{fig:cwb1_3d_sliceyz_montage}
and~\ref{fig:cwb1_3d_slicexzxy_montage} show various slices through
the simulation of model cwb1, which allow one to more fully appreciate
the complex 3D structure of the winds and their collision.  In systems
with equal wind momenta, like model cwb1, the WCR twists like a helix
out of the orbital plane. The WCR thickens above and below the orbital
plane, and its orientation increasingly lags that within the orbital
plane as one moves further from the orbital plane. The dense gas
remains clumpy and fragmented, and is gradually accelerated as it
moves out of the orbital plane.

Fig.~\ref{fig:statistics}(a) shows histograms of the mass within
$120\,\Rsol$ of the system centre of mass, divided into density bins
of width 0.1 in log. Gas at all temperatures is included, but the
densest gas within 1.2 stellar radii is excluded. The highest density
then obtained is $2.4\times10^{-12}\,{\rm g\;cm^{-3}}$, which occurs
in the cold postshock gas between the stars. The mass as a function of
density rises rapidly as the density decreases until
$\rho\approx2\times10^{-14}\,{\rm g\;cm^{-3}}$, after which it is
roughly constant until the density is approximately 2.5 dex lower.  A
low density tail extends to $\rho\approx2\times10^{-18}\,{\rm
g\;cm^{-3}}$.  Such rarefied gas exists in downstream regions of the
WCR which are ``shadowed'' by the cool dense ribbon of gas closer to
the WCR apex (cf. Fig.~\ref{fig:cwb123_density}a).

To better understand Fig.~\ref{fig:statistics}(a) we also plot the
mass distribution between radii of $12-120\,\Rsol$ from a single O6
star.  If material leaves the surface of the star at terminal velocity
the slope of the resulting mass distribution (labelled as ``vinf'' in
Fig.~\ref{fig:statistics}a) is $-0.5$. An accelerating wind induces
curvature into this slope, such that there is relatively more mass at
higher densities (this distribution is labelled as ``CAK'' in
Fig.~\ref{fig:statistics}a). Comparing the ``CAK'' and ``cwb1''
profiles we see that the overall effect of a WCR is to create much
more higher density gas in the immediate circumstellar environment
than would otherwise be the case.

Fig.~\ref{fig:statistics}(b) shows the mass within radii
$12-120\,\Rsol$ as a function of speed, in bins of $50\;\kmps$ width.
The ``CAK'' profile shows that most of the mass in the single-star
case has been accelerated to nearly the terminal wind speed. In
contrast, in model cwb1 the distribution is reasonably flat between
$1000-3000\kmps$, with less mass at lower speeds.
Fig.~\ref{fig:statistics}(b) also reinforces the point that there is
significant mass moving at speeds up to $500\kmps$ faster than the
terminal speeds of the individual stars, while a smaller proportion of
mass exceeds the terminal wind speed by $\sim 1000\;\kmps$.

Fig.~\ref{fig:statistics}(c) shows the mass in model cwb1 as a 
function of temperature, in bins of width 0.1 in log. 
Most of the mass (unshocked plus shocked
material) is cold ($\approx 10^{4}\,K$). Some material is shock
heated above $10^{7}\,$K, with the majority of the hot ($T > 10^{5}$\,K)
gas at temperatures of a few million degrees. Note that the mass
distribution between a few times $10^{4}$\,K and a few times $10^{7}$\,K 
bears some resemblance to an inverted ``cooling curve'', with less
gas at temperatures where cooling is very rapid. This behaviour is
quite different to the mass distributions of adiabatic wind-wind
collisions plotted in \citet{Lemaster:2007}.

The volume of gas as a function of temperature is shown in
Fig.~\ref{fig:statistics}(d). Most of the volume within the
investigated region contains cold gas.

\subsection{Model cwb2}
\label{sec:dynamics_cwb2}
A density snapshot of model cwb2 is shown in
Fig.~\ref{fig:cwb123_density}(b) where it can be seen that the
character of the WCR is completely different to that of model cwb1.
The most striking difference is that the postshock gas remains largely
adiabatic, flowing out of the system while still relatively hot. This
was anticipated since the estimated value of $\chi$ is substantially
larger than unity. The greater distance between the stars allows the
winds to accelerate along the line-of-centres to $\approx 1630 \kmps$
before their collision (c.f. a speed of $1820 \kmps$ at the same
distance in the single star case). This results in postshock
temperatures up to $\approx 4 \times 10^{7}\;$K
(Fig.~\ref{fig:statistics}g), while the maximum postshock density is
$\approx 4 \times 10^{-15} \gpcm3$ (Fig.~\ref{fig:statistics}a),
compared to $\approx 2.4\times10^{-12}\gpcm3$ in the cool, dense
postshock gas in model cwb1.

The predicted aberration angle is $\approx 7^{\circ}$. This is again
higher than is observed ($\approx 3-4^{\circ}$) for the same reasons
as mentioned for model cwb1.  Fig.~\ref{fig:cwb123_density}(b) also
shows that there is a density gradient across the flow direction in
the downstream parts of the WCR (i.e. there is a reduction in density
from the contact discontinuity towards the leading shock, and an
increase towards the trailing shock). This is similar to the behaviour
already noted in model cwb1, being caused by the slower exit out of
the system of the gas in the WCR than in the winds, and the shadowing
introduced by the curvature of the WCR which results in low densities
ahead of the leading shock, and therefore seems to be a general
property of CWBs with equal winds (the unequal winds case, discussed
in Sec.~\ref{sec:dynamics_cwb3}, is more complicated).

The typical speed of gas in the WCR as it leaves the hydrodynamical
grid in the orbital plane as shown in Fig.~\ref{fig:cwb123_density}(b) 
is about $2500\;\kmps$. In 10 days this gas travels a distance of
$14.4\;{\rm au}$ ($3100\;\Rsol$). The minimum distance from the centre of the
grid to its edge is $285\,\Rsol$, so we expect the WCR to curve
through 9 per cent of a full rotation, or about $33^{\circ}$.
At a distance of $285\,\Rsol$ from the centre of mass, the WCR in
Fig.~\ref{fig:cwb123_density}(b) actually displays a curvature
of just over $60^{\circ}$, again indicating that the speed of gas
near the apex of the WCR is much slower than further downstream.
Hence the shape of the WCR near its apex is not very well approximated
by an Archimedean spiral since the speed within the WCR is not constant.

We again overplot results from the dynamical model of \citet{Parkin:2008}.
The contact discontinuity traced by the red line was calculated 
assuming terminal wind speeds of $2500\;\kmps$, while the green line
is from a model with wind speeds of $1650\;\kmps$. Both models assumed
that the shocked wind flowed ballistically once it had reached 85 per cent of
its terminal speed. Clearly the model which assumes the higher terminal wind
speeds is a better match to the hydrodynamical calculation, which
indicates that the downstream morphology of the WCR is not solely
dependent on the pre-shock wind speeds along the line of centres.

\citet{Lemaster:2007} noted that due to the curvature of the WCR by
the orbital motion of the stars, KH instabilities occurred even when
the winds had equal speeds. There is no sign of this in our model, but
this could be because of the higher ratios of $v_{\rm orb}/v_{\rm w}$
in some of Lemaster et al.'s models. The fastest wind speed attained
in model cwb2 in the orbital plane is $\approx 3200 \kmps$, near the
trailing shocks of the WCR as in model cwb1. The lowest wind speeds in
the orbital plane occur in the top left and bottom right of the panel
shown in Fig.~\ref{fig:cwb123_velocity}(b), where the speed is limited
to $\approx 2400 \kmps$. Both these values are lower than the
corresponding values obtained from model cwb1, indicating that the
boosts to the acceleration of certain regions of the winds are greater
the closer the two stars (and the centres of their radiation fields)
are to each other.

Fig.~\ref{fig:cwb123_temp}(b) displays the temperature of the WCR in
the orbital plane.  The gas near the trailing shocks cools to $\approx
5 \times 10^{6}\;$K as it reaches the edge of the grid, but close to
the leading shocks the postshock temperature is $\approx 7 \times
10^{5}\;$K near the edge of the grid. The lower temperatures near the
leading shocks arise because the leading shocks are more oblique at a
given downstream distance than the trailing shock. Hence, in systems
with identical winds the gas near the leading shocks is both less
dense and cooler than gas near the trailing shocks, and
vice-versa. The situation changes slightly in systems with unequal
winds, as noted in Section~\ref{sec:dynamics_cwb3}.

In model cwb2 the postshock ion and electron temperatures
at the apex of the WCR equilibrate in about 2 hours. This is
about an order of magnitude faster than the flow time of this gas out
of the system ($t_{\rm flow} \sim (D_{\rm sep}/2)/(v_{\rm w}/4) \sim 7
\times 10^{4}\;{\rm s}$). Further downstream there is a thin postshock
region where there is a noticeable delay for the electrons to heat to
the same temperature as the ions. However, the thickness of these
regions is typically only a few per cent of the width of the WCR at
these positions. Given this and the reduced plasma emissivity compared
to the apex of the WCR, it is not expected that there will be any
observable effects revealing $T_{\rm e} < T_{\rm ion}$ in systems
similar to this model. However, slow electron heating is certainly
important in wider systems, such as WR\,140, as other work has already
pointed out \citep{Zhekov:2000,Pollock:2005}.

The ionization age, $n_{\rm e}t$, is $ \approx 2 \times 10^{12} \;{\rm
cm^{-3}\;s}$ in the postshock gas near the apex of the WCR, and
reaches $4 \times 10^{13} \;{\rm cm^{-3}\;s}$ at the contact
discontinuity in the densest part of the WCR (i.e. near the stagnation
point). Further downstream the postshock flow does not quickly
reach ionization equilibrium. However, the central part of the WCR
(close to the contact discontinuity) remains in collisional ionization
equilibrium, as this gas originates from close to the WCR apex.

Figs.~\ref{fig:cwb2_3d_sliceyz_montage}
and~\ref{fig:cwb2_3d_slicexzxy_montage} show various slices through
the simulation of model cwb2, again to allow one to more fully
appreciate the 3D structure of the winds and their collision.
Compared to model cwb1, we once again see that the interaction region
is much smoother and more stable. The top right panel of
Fig.~\ref{fig:cwb2_3d_slicexzxy_montage} reveals that the winds
accelerate faster out of the orbital plane than regions within it
where one of the stars shadows the other, due to the more effective
combination of the stellar radiation fields.

Fig.~\ref{fig:statistics}(a) shows a histogram of mass versus density
within $120\,\Rsol$ of the centre of mass of model cwb2 (excluding material
within $14\,\Rsol$ of the centres of the stars).
The distribution outlined in blue shows that most of the mass is
at a density of $\approx 5 \times 10^{-16}\,{\rm g\, cm^{-3}}$,
and extends to $\approx 4 \times 10^{-17}\,{\rm g\, cm^{-3}}$ at the
low density end and $\approx 6 \times 10^{-14}\,{\rm g\, cm^{-3}}$
at the high density end, the latter representing unshocked wind material
close to the stars. Compared to model cwb1 the distribution is
not as flat or as wide.

Fig.~\ref{fig:statistics}(a) also shows the mass-density distribution
for only the hot ($T > 10^{5}$\,K) gas. The maximum density is now
$\approx 4 \times 10^{-15}\,{\rm g\, cm^{-3}}$. The full (blue)
distribution is a rough combination of two ``CAK'' distributions (one
for each star) plus the hot (pink) distribution.  However, it is clear
that the sum of these distributions would lead to more mass at
densities below $\approx 2 \times 10^{-16}\,{\rm g\, cm^{-3}}$ than is
actually seen in model cwb2. The excess mass at higher densities in
model cwb2 is because the WCR gas flowing through the spherical
surface of radius $120\Rsol$ has $\rho \gtsimm 2 \times 10^{-16}\,{\rm
g\, cm^{-3}}$.

The mass-speed distribution is shown in Fig.~\ref{fig:statistics}(b).
Compared to model cwb1 we find a reduced maximum speed attained by the
gas (with the distribution turning over at $\approx v_{\infty}$, rather 
than at $\approx 500\kmps$ above the single
star terminal wind speed). There is also less gas moving at speeds below
$1250\kmps$, and more mass moving at speeds between $1250-2600\kmps$.
Limiting our analysis to only hot ($T > 10^{5}$\,K) gas we see that there
is a peak in the distribution near $1500\kmps$. The maximum speed of hot
gas in the orbital plane is $2500\kmps$, increasing to nearly
$3000\kmps$ out of the orbital plane.

Fig.~\ref{fig:statistics}(c) shows the distribution of mass versus
temperature. The majority of the gas mass is predominantly cold (since
the gas in the WCR stays hot in model cwb2 as it flows out of the
system, the cold material represents the unshocked winds). However,
there is also significant mass at temperatures $\sim 10^{7}$\,K. The
amount of mass at intermediate temperatures ($\sim 10^{4} - 7 \times
10^{5}$\,K) should be treated cautiously, since it mostly represents
gas in cells at the unresolved shocks, and will therefore be sensitive
to the numerical scheme.  The bi-modal nature of the mass-temperature
distribution of model cwb2 is also reflected in
Fig.~\ref{fig:statistics}(d), which shows the volume-temperature
distribution. 76 per cent of the volume within $120\,\Rsol$ of the
system centre of mass contains the unshocked winds, while 20 per cent
contains gas at temperatures exceeding $5 \times 10^{6}$\,K.

Figs.~\ref{fig:statistics}(e)-(h) show mass and volume distributions
of the hot gas in model cwb2 out to a radius of $285\,\Rsol$ from the
system centre of mass. The distributions are similar to those in
Figs.~\ref{fig:statistics}(a)-(d), but extend to lower densities and
temperatures, and greater speeds, as expected.
 
\subsection{Model cwb3}
\label{sec:dynamics_cwb3}
A distinct difference between the WCR in model cwb3 compared to the
other models is that its apex occurs closer to one star (the
O8V secondary star which has the weaker wind) than the other star (the
O6V primary star) due to the unequal wind momentum ratio (see
Fig.~\ref{fig:cwb123_density}c). The gas
downstream of the WCR apex is also pushed closer to the
secondary star, due to the greater momentum flux of the primary
wind. These effects, together with the fact that the winds are still
accelerating when they collide, plus the enhanced terminal wind speed
of the O6V star, lead to higher preshock and postshock speeds in the
primary wind relative to the secondary wind. This has three major
consequences: i) there is a velocity shear along the contact
discontinuity between the winds, which excites Kelvin-Helmholtz
instabilities; ii) to maintain pressure balance, the preshock and
postshock density of the secondary wind exceeds that of the primary
wind; iii) higher postshock densities, together with reduced postshock
speeds, results in the gas on the secondary side of the WCR radiating
more efficiently (this is discussed in more detail in a forthcoming
paper on the X-ray emission).
 
There is also a clear signature of radiative inhibition: at the
primary wind shock on the line-of-centers, the pre-shock speed is
$1800\;\kmps$, whereas without the companion star it is $2010\;\kmps$
at this distance.  Hence the radiation field of the secondary star
reduces the net acceleration of the primary wind towards it by
$\approx 10$\,per cent. Likewise, the pre-shock speed of the secondary
wind on the line-of-centers is $1270\,\kmps$, whereas the single star
solution has a speed of $1450\;\kmps$ at this distance, indicating a
reduction in the acceleration of $\approx 12$\,per cent. Radiative
inhibition thus has a greater effect on the secondary wind, as
expected. There is no sign of radiative braking \citep{Gayley:1997}, 
though this is not surprising given the low wind momentum ratio (hence
wind opacity ratio) in our model.

Also overplotted on Fig.~\ref{fig:cwb123_velocity}(c) is the
comparison with the dynamical model of \citet{Parkin:2008}. The match is
again very good, with the main differences between the models
occuring in the leading arm where the contact discontinuty in the
hydrodynamical model shows slightly greater curvature.

Fig.~\ref{fig:cwb123_velocity}(c) shows the speed of gas in the
orbital plane. The maximum speeds reached in the O6 and O8 winds are
$2910\;\kmps$ and $2840\;\kmps$, respectively. These values are
attained near the trailing edges of the WCR at the boundaries of our
hydrodynamic grid. More typically, the O6 and O8 winds reach speeds of
$\approx 2550\;\kmps$ and $2300\;\kmps$ (again in directions towards
the top left and bottom right of the hydrodynamic grid), respectively.
Gas in the denser parts of the WCR typically reaches speeds of
$1600\kmps$ in the orbital plane by the time it leaves the
hydrodynamic grid.
  
Using the measured pre-shock velocities along the line of centres to
determine the cooling parameter $\chi$, we find that $\chi_{1} \approx
28$ and $\chi_{2} \approx 14$.  Thus the postshock winds remain
largely adiabatic, as is indeed seen.  The postshock winds reach
maximum temperatures of $\approx 5\times 10^{7}\;$K (for the shocked
O6 wind) and $\approx 3\times 10^{7}\;$K (for the shocked O8 wind),
in good agreement with theoretical expectations. The maximum densities
attained are $\approx 2\times10^{-15}\gpcm3$ and $5\times10^{-15}\gpcm3$ for
the shocked O6 and O8 winds, respectively.  These maximum values are
all attained near the stagnation point.  

\begin{figure*}
\psfig{figure=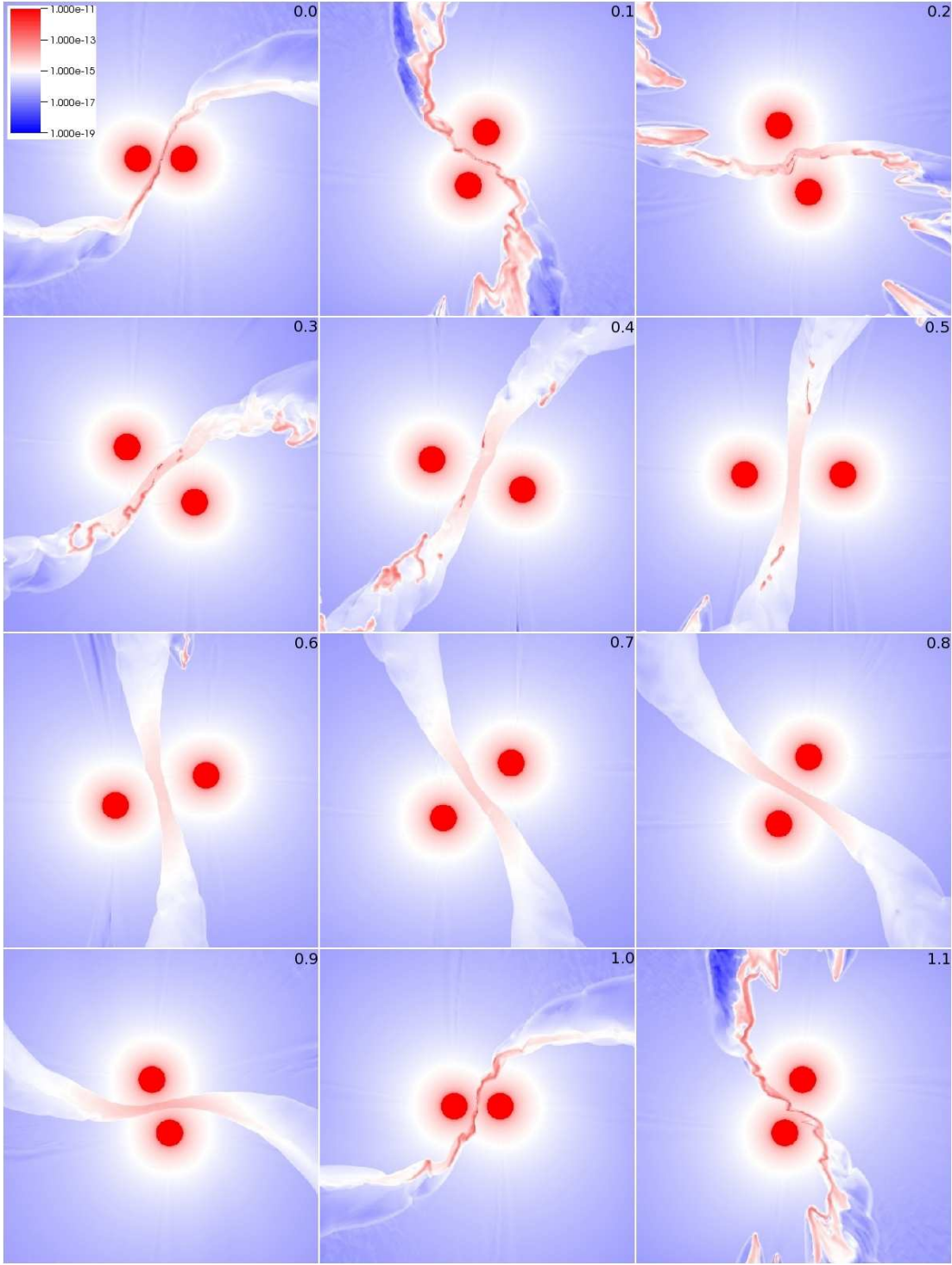,width=14.6cm}
\caption[]{Density snapshots of model cwb4 at specific phases in the
orbit, which progresses from left to right, and from top to
bottom. The top left panel shows the model at periastron, and
subsequent panels increase the orbital phase in steps of 0.1. The
right most panel of the 2nd row from the top shows the stars at
apastron, with periasron again occuring in the middle panel of the
bottom row. The colour scale is logarithmic, spanning $10^{-19}\gpcm3$
(blue) to $10^{-11}\gpcm3$ (red). The panels have sides of length
$240\;\Rsol$.}
\label{fig:cwb4_density}
\end{figure*}

\begin{figure*}
\psfig{figure=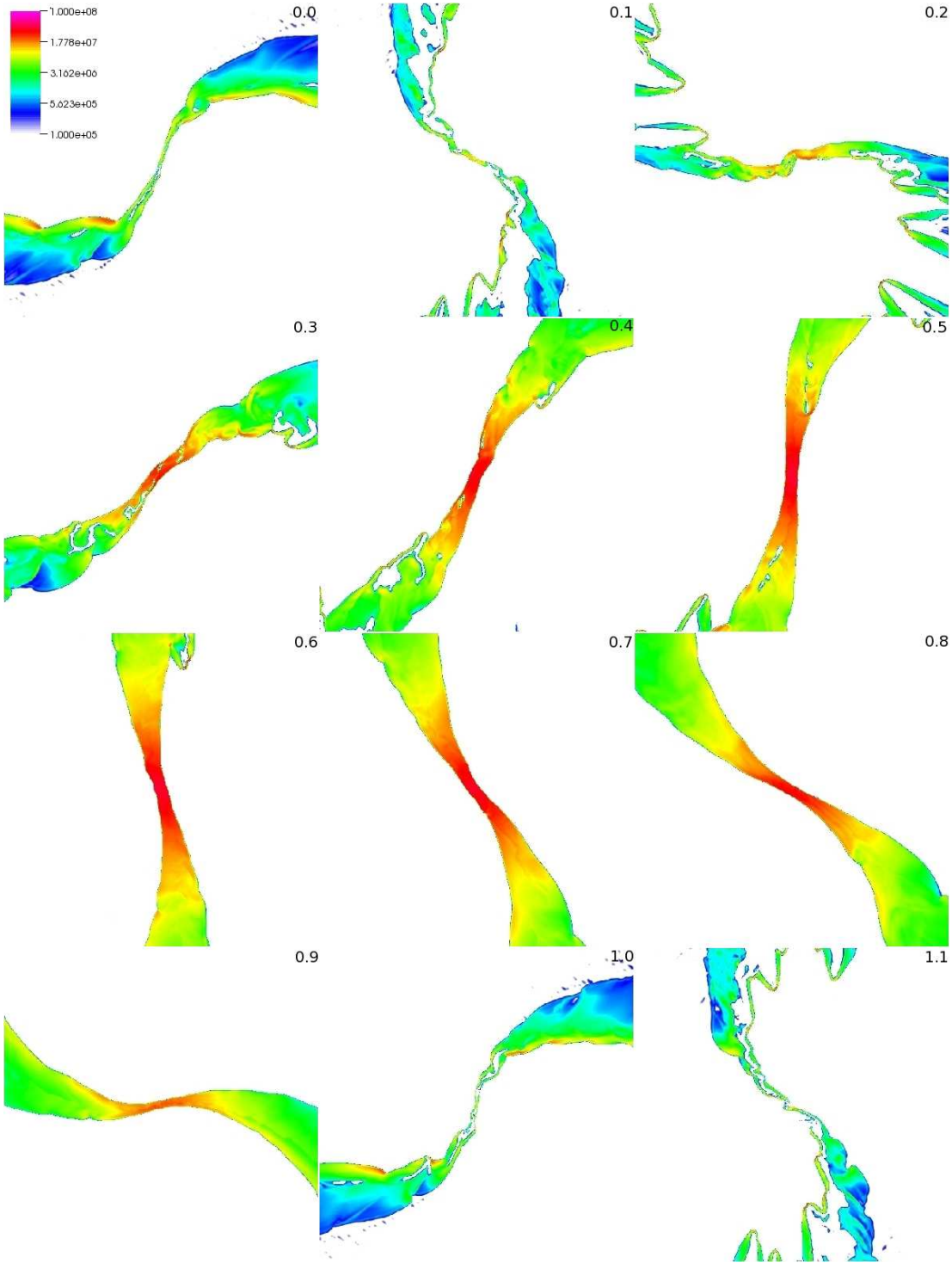,width=14.6cm}
\caption[]{As Fig.~\ref{fig:cwb4_density} but showing temperature snapshots. 
The colour scale is logarithmic, spanning $\leq 10^{5}$\,K
(white) to $\geq 10^{8}$\,K (pink).}
\label{fig:cwb4_temperature}
\end{figure*}

\begin{figure*}
\psfig{figure=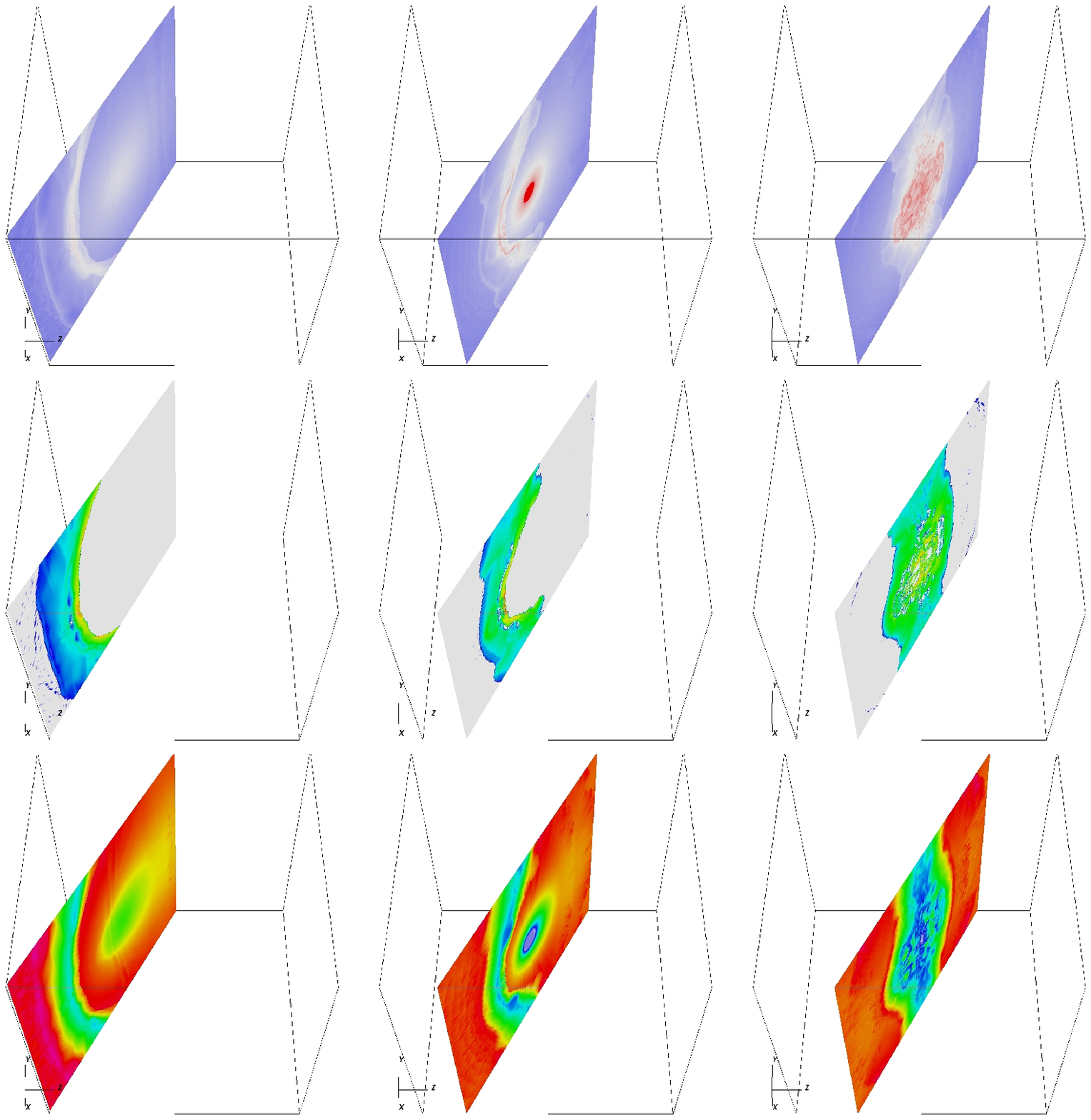,width=15.2cm}
\caption[]{Density (top), temperature (middle), and speed (bottom)
slices of model cwb4 parallel to the surface of the WCR near its apex
at periastron. The left panels pass through the WCR downstream of its
apex, the middle panels pass through the centre of one of the stars,
and the right panels pass through the stagnation point and apex of the
WCR. The colour scales are the same as in
Fig.~\ref{fig:cwb1_3d_sliceyz_montage}.}
\label{fig:cwb4_3d_slice_parallelwcr_montage}
\end{figure*}

The ratio of post-shock densities across the contact discontinuity
(CD) is found to vary with downstream distance. The factor of 2.5
difference at the stagnation point increases to a factor of $7-10$ in
the arm where the O8 gas forms the trailing edge, but becomes a factor
of 3 or so in favour of the O6 wind (i.e. the shocked O6 material is
denser than the shocked O8 material) in the opposite arm.  The
temperature difference across the CD increases in the WCR arm where
the O6 material is at the trailing edge, whereas it inverts (i.e. the
shocked O8 material becomes hotter, through less rapid cooling, than
the O6 material) in the opposite arm.

The postshock electron and ion temperatures rapidly attain equilibrium
in all parts of the flow, in common with the previous models. The
significantly higher density and lower temperature of the shocked O8
wind at the stagnation point leads to more rapid ionization
equilibrium compared to the shocked O6 wind. At the WCR apex, the
ionization age of the shocked O6 wind is typically $4.5\times10^{12}
\;{\rm cm^{-3}\;s}$, while nearer to the CD it increases to $\approx
3.5\times10^{13} \;{\rm cm^{-3}\;s}$. The ionization age of gas near
the CD further increases downstream, as already noted for the other
models.
  
Fig.~\ref{fig:statistics}(e) shows the mass-density distribution of
hot gas ($T > 10^{5}$\,K) in model cwb3 within a spherical volume
centered on the system centre of mass and extending to a radius of
$285\,\Rsol$ (again excluding material within $14\,\Rsol$ of the
centres of the stars). The distribution is similar to that from model
cwb2 over an identical volume, though the total mass is somewhat less
owing to the smaller mass-loss rate of the secondary star in model
cwb5. However, it is interesting to find that the mass at higher
densities is dominated by material from the O8 wind, and vice-versa.
This is because the shocked O8 wind is denser at the WCR apex than the
O6 wind (due to the latter's greater pre-shock speed), and remains
denser as it flows out of the system due to the curvature/confinement
of the O8 gas as the WCR bends behind the O8 star.

The mass-speed distribution is shown in Fig.~\ref{fig:statistics}(f).
Compared to the distribution obtained from model cwb2, the cwb3
distribution extends to a slower maximum speed. This is due to the
lower terminal velocity of the O8 wind, and the reduction in the
combined radiative driving force due to the reduced luminosity of the
O8 star.  Model cwb3 has more mass at speeds $< 1400 \kmps$ - this is
dominated by material from the O8 star, and indicates the slower
acceleration of its wind, and subsequently the lower speed of its
postshock gas. Conversely, material from the O6 star dominates the hot
gas moving at speeds above $1600 \kmps$. These differences may have
implications for the emission line profiles from the WCR of such
systems.

Fig.~\ref{fig:statistics}(g) shows the mass-temperature distribution.
Compared to model cwb2, the distribution from model cwb3 contains less
mass overall (as already noted), but also has gas at a higher maximum
temperature, since the O6 wind has more room to accelerate along the
line-of-centres before being shocked. Model cwb3 also contains much
more mass at $T < 2\times10^{6}$\,K because of the slower acceleration
and speed of the O8 wind.

The volume-temperature distribution of hot gas ($T > 10^{5}$\,K) is
shown in Fig.~\ref{fig:statistics}(h).  Compared to model cwb2 we find
that the volume of hot gas above $4 \times 10^{7}$\,K is greater in
model cwb3, due to the faster pre-shock speed of the O6 wind. The
volume of hot gas is slightly less between $4 \times 10^{6} - 4 \times
10^{7}$\,K, is comparable at $\approx 3 \times 10^{6}$\,K, and is
greater for $T < 1.4 \times 10^{6}$\,K.  Shocked O6 gas dominates the
volume of hot plasma at $T > 2 \times 10^{6}$\,K, below which the O8
wind dominates.

The emission measure ($n_{\rm e}^{2}V$) of the hot gas, divided into
temperature bins of width 0.1 in log, peaks at $9.2\times10^{54}\;{\rm
cm^{-3}}$ at $T\approx1.75\times10^{7}$\,K (this is not shown in
Fig.~\ref{fig:statistics}).  The shocked O8 wind is
almost completely dominant, with the emission measure from O6 wind
material peaking at $2.4\times10^{54}\;{\rm cm^{-3}}$ at the slightly
hotter temperature of $T\approx2.25\times10^{7}$\,K. Thus the shocked
O8 wind should dominate, for example, the thermal radio and X-ray
emission generated in the WCR - in fact, we find that the O8 wind
dominates the intrinsic X-ray emission below $4.5$\,keV 
\citep[][in preparation]{Pittard:2009}. The total emission measure is
$7.1\times10^{55}\;{\rm cm^{-3}}$, with the shocked O8 wind contributing
70 per cent of this value. Since the shocked O8 material in the
leading arm of the WCR is compressed into about half of the volume
of the shocked O6 material in this arm (note the high density
contrast in Fig.~\ref{fig:cwb123_density}c), it is interesting to 
examine the contributions of each wind to the emission measure from each arm
of the WCR. We find that the shocked O8 material dominates the emission
measure in the leading arm ($2.8\times10^{55}\;{\rm cm^{-3}}$ versus
$0.8\times10^{55}\;{\rm cm^{-3}}$ for the shocked O6 wind). In the trailing
arm the ratio of the emission measures between the winds is much closer
to unity ($2.0\times10^{55}\;{\rm cm^{-3}}$ for the shocked O8 wind, versus
$1.4\times10^{55}\;{\rm cm^{-3}}$ for the shocked O6 wind), though the
O8 material again dominates the emission.

\subsection{Model cwb4}
\label{sec:dynamics_cwb4}
As one might expect given the large differences between models cwb1
and cwb2, the structure and properties of the WCR in model cwb4
undergoes rich and dramatic changes as a result of the varying
separation between the stars as they progress in their respective
orbits.  Fig.~\ref{fig:cwb4_density} shows the evolution of the mass
density in the plane of the orbit. At periastron the WCR is highly
radiative and subject to strong dynamical instabilities. The cooled
postshock gas is sheet-like, as in model cwb1, and maintains a
coheasive structure. As the stars separate and the cooling of the
postshock gas becomes less severe, this sheet fragments and
discombobulates into smaller, individual clumps.  By apastron, the WCR
is largely adiabatic. At both periastron, when the stars are separated
by the same distance as in model cwb1, and at apastron, when the stars
are separated by the same distance as in model cwb2, the wind-wind
interaction somewhat resembles the WCR in these other models (cf also
Fig.~\ref{fig:statisticscwb4}).  For instance, the velocity profiles
between the stars along the line of centres from model cwb4 at
periastron and apastron are in good agreement with those from models
cwb1 and cwb2, respectively.

Nevertheless, there are some interesting differences. For instance,
some dense clumps of cool gas exist within the WCR even at apastron,
in contrast to the entirely hot state of the WCR in model cwb2. These
clumps were formed during the previous periastron passage. Because of
their high density and inertia relative to the surrounding hot, more
rarefied, plasma, it can take some considerable time for the faster
flowing hot plasma to push the clumps out of the system. Clearly this
timescale is of the order of half the orbital period of the stars in
model cwb4. The slow acceleration of the clumps out of the system also
results in some of the clumps crossing the global shocks bounding the
WCR and moving into the fierce high speed environment of the unshocked
winds (this is clearly seen at phases $0.4-0.6$ in
Fig.~\ref{fig:cwb4_density}).  This process occurs because the hot
plasma in the WCR responds quite quickly to the movement of the stars
(and thus rotates with the stars), whereas the direction vectors of
the dense clumps respond much less rapidly and the clumps tend to
maintain straighter paths. For this reason the clumps exit only out of
the trailing shocks of each arm of the WCR, and not out of the leading
shock. Once formed, the clumps are gradually destroyed by thermal
evaporation and ablation, with larger and denser clumps surviving
longer.

The presence of clumps in or around an otherwise adiabatic WCR may
affect some observational signatures. For instance, gas mixed from the
clumps into the surrounding flow ``mass-loads'' it, making it
denser. The effect of the clump on the temperature of the surrounding
flow is more complex. A bowshock forms upstream when the surrounding
flow is supersonic with respect to the clump, heating the surrounding
flow. This is seen, for example, when the clumps move out of the WCR
and into the unshocked winds. Bowshocks also form upstream of clumps
{\em within} the WCR when the clumps are far enough downstream that
the hotter flow past them is supersonic - this is the case for the
lowermost clump above the WCR apex in the apastron panel of
Fig.~\ref{fig:cwb4_density}. Friction between the clump and a
surrounding subsonic flow also heats the flow. Further downstream, the
addition of mass into a flow generally leads to lower temperatures
once the injected mass is fully mixed into the flow
\citep[e.g.][]{Pittard:2003a}, though this will require reconnection of
field lines in a magnetic environment.  The presence of clumps outside
the main WCR may also lead to more of the winds' kinetic power being
processed through shocks than would otherwise be the case, and may
affect the variation of, for example, the X-ray emission with
phase. Finally, the boundary layer between clumps and the surrounding
diffuse flow contains intermediate densities, temperatures, and
velocities, all of which affect the resulting emission
\citep[][]{Hartquist:1993}.  Line emission is expected to be
particularly sensitive to the conditions in this region, and such
regions may help to explain discrepancies in current models
\citep[e.g.][]{Henley:2008} where mixing of cold material into a
hotter medium is neglected.

Another difference with our previous models is that cold dense gas
does not exist far downstream of the apex of the WCR when the stars
are at periastron, whereas it does in model cwb1 (compare the top left
panels in Figs.~\ref{fig:cwb4_density} and~\ref{fig:cwb4_temperature}
with Figs.~\ref{fig:cwb123_density}(a) and~\ref{fig:cwb123_temp}(a),
and remember that the stellar separations are identical). This is
simply because the downstream gas in model cwb4 at this phase reflects
the less radiative environment which existed within the WCR when the
stars were more widely separated.

Not surprisingly, there is a large variation in the aberration and
curvature of the WCR in model cwb4 with orbital phase. The aberration
angle reaches a maximum at periastron due to increased stellar
velocities and reduced preshock wind speeds. At periastron,
$v_{\rm orb} = 334\,\kmps$, while $v_{\rm w} \approx 710\,\kmps$,
giving an expected aberration angle of $\approx 25^{\circ}$, in good
agreement with the measured angle of $\approx 21^{\circ}$. At
apastron, $v_{\rm orb}$ drops to $156\,\kmps$, while $v_{\rm w}$
increases to $\approx 1665\,\kmps$, giving an expected aberration
angle of $\approx 5^{\circ}$, again in good agreement with the
measured angle of $\approx 4^{\circ}$. Note that the orbital speeds at
periastron and apastron are different to the speeds in models cwb1 and
cwb2, so the aberration and curvature of the WCR differs
compared to the circular orbit models (see also Table~\ref{tab:keyparams}). 

Fig.~\ref{fig:cwb4_temperature} shows the temperature in the orbital
plane of model cwb4.  The hottest gas occurs at apastron ($T_{\rm max}
\approx 5 \times 10^{7}\;$K), and its temperature is similar to the
hottest gas in model cwb2. While the maximum gas temperature attains a
minimum around periastron ($T_{\rm max} \approx 2 \times 10^{7}\;$K),
this value is found in localized bowshocks enveloping dense clumps in
the downstream flow and is sensitive to rapid changes in the local
dynamics. It also relates to a very small fraction of the total mass
within the WCR, and hence is not representative of the typical
temperatures which are much lower
(cf. Fig.~\ref{fig:statisticscwb4}e).
Fig.~\ref{fig:cwb4_temperature} also reveals that at periaston the
trailing edge of each arm of the WCR is hotter than the leading
edge. This is because of the severe curvature of the WCR due to
orbital motion, which results in the shock at the trailing edge
staying more normal to the preshock wind vector. This effect is not
seen at apastron when the effects of orbital motion are less intense
(see Fig.~\ref{fig:cwb4_temperature}).  However, model cwb2 suggests
that gas at the trailing edge of each arm of the WCR will become
hotter than gas at the leading edge at distances further downstream
than can be probed in model cwb4 (model cwb2 has a larger grid - see
Table~\ref{tab:grid}).

\begin{figure}
\psfig{figure=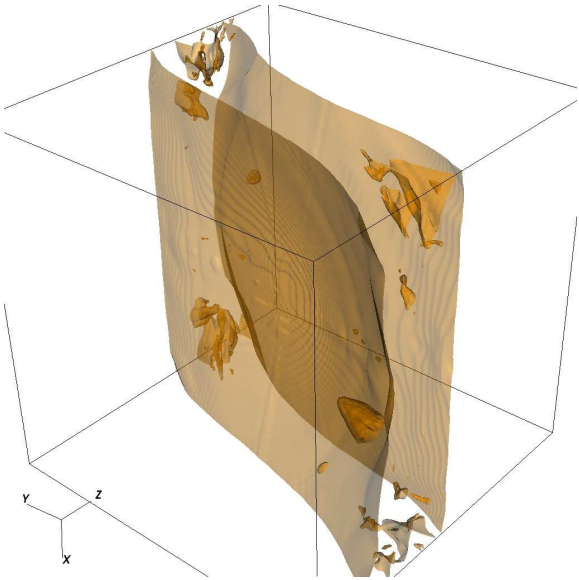,width=7.6cm}
\caption[]{3D temperature surface from model cwb4 at apastron. 
The surface traces gas at a temperature of $3\times10^{6}$\,K.}
\label{fig:cwb4_3ddensity}
\end{figure}

Fig.~\ref{fig:cwb4_3d_slice_parallelwcr_montage} shows various slices
through the 3D hydrodyanmical grid when the stars are at periastron,
again to allow appreciation of the 3D structure of the winds and their
collision.  The slices in the right-most panels pass through the
stagnation point at an angle aligned with the shock surfaces between
the stars, and so sample a relatively large distance through the WCR
despite its small width.  This figure illustrates again that cold
dense gas exists near the apex of the WCR at periastron, but not
further downstream at this phase.

\begin{figure*}
\psfig{figure=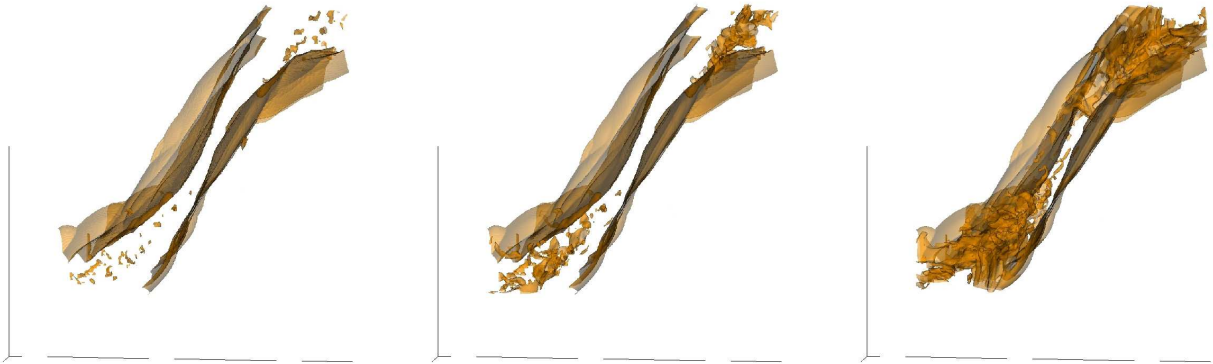,width=17.1cm}
\caption[]{3D temperature surface from model cwb4 at $\phi=0.4$. The surface
traces gas at a temperature of $2\times10^{5}$\,K (left),
$10^{6}$\,K (middle), and $3\times10^{6}$\,K (right).}
\label{fig:cwb4_3dtemp2}
\end{figure*}

\begin{figure*}
\psfig{figure=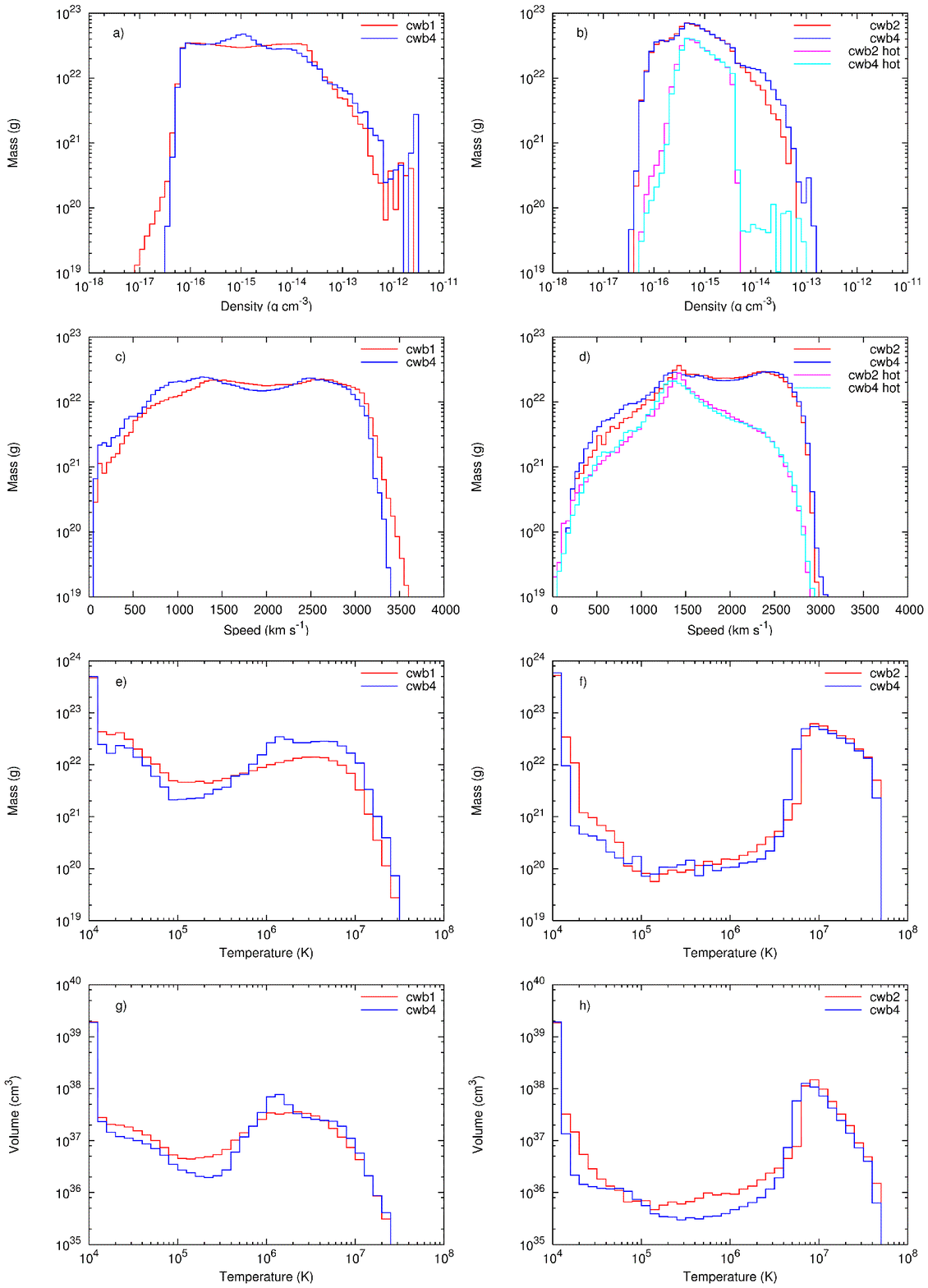,width=14cm}
\caption[]{Histograms of various quantities in model cwb4, compared
to models cwb1 and cwb2. The left
column shows data from models cwb1 and cwb4 with the later at periastron,
while the right column compares model cwb2 with cwb4 at apastron.
All histograms examine the material within a radius of
$120\;\Rsol$ of the system centre of mass, and exterior to 1.2 stellar
radii around the stars.}
\label{fig:statisticscwb4}
\end{figure*}

Fig.~\ref{fig:cwb4_3ddensity} displays a 3D temperature surface from
model cwb4 at apastron. The sheet-like structures of the global shocks
bounding the WCR are clearly visible. However, this figure also
highlights that while cold gas has been largely cleared out of the
central regions of the WCR, some still remain further downstream. The
large bowshock projected towards the bottom right of the plot
envelopes the dense blob of gas seen outside of the WCR towards the
bottom left of the $\phi=0.5$ panel in
Fig.~\ref{fig:cwb4_density}. Fig.~\ref{fig:cwb4_3dtemp2} shows
temperature surfaces viewed from above the orbital plane at phase
$\phi=0.4$. Fig.~\ref{fig:cwb4_3dtemp2}(a) traces gas at a temperature
of $2\times10^{5}$\,K, which highlights the global shocks bounding the
WCR and the interface between cold, dense clumps embedded within the
hotter, more rarefied plasma of the WCR.
Figs.~\ref{fig:cwb4_3dtemp2}(b) and~(c) show surfaces of progressively
higher temperatures. As the temperature of these surfaces increases
the intermediate-temperature material ablated from the cold clumps is
revealed. Comparison between Figs.~\ref{fig:cwb4_3ddensity}
and~\ref{fig:cwb4_3dtemp2}(c) highlights the gradual removal of cold
dense clumps from the WCR as the orbit progresses away from the
previous periastron passage.

Fig.~\ref{fig:statisticscwb4}(a) shows the mass within $120\,\Rsol$ of
the system centre of mass (excluding the stars) as a function of
density from model cwb4 at periastron, and from model cwb1. The
distributions are reasonably similar, with the main differences being
a relative excess of mass at high ($\rho > 5 \times 10^{-14}\,{\rm g
\,cm^{-3}}$) and intermediate ($\rho \sim 10^{-15}\,{\rm g
\,cm^{-3}}$) densities, and a lack of low density material ($\rho < 3
\times 10^{-16}\,{\rm g \,cm^{-3}}$) in model cwb4. The first of these
is explained by the recent history of the interaction in model cwb4 -
i.e.  that there is hotter gas in the downstream regions of the WCR
because this gas was shocked and advected out of the central regions
of the system before cooling became as strong as it is now that the
stars are at periastron. The last is explained by a reduction in the
degree of ``shadowing'', since prior to periastron the rate at which
the stars sweep out angle is much smaller than the rate in model cwb1.

A comparison between the mass versus density histograms of models cwb2
and cwb4 at apastron (Fig.~\ref{fig:statisticscwb4}b) again reveals
similar distributions, with the most noticeable difference being a
relative excess of mass at $\rho > 5 \times 10^{-15}\,{\rm g cm^{-3}}$
in model cwb4.  This now reflects the presence of cold dense clumps in
the WCR from the previous periastron passage.  A comparison of the
mass of hot ($T > 10^{5}\,$K) gas reveals a similar excess at higher
densities, due to the intermediate temperatures of gas ablated off the
cold dense clumps.

The mass versus speed histogram of models cwb1 and cwb4 at periastron
are shown in Fig.~\ref{fig:statisticscwb4}(c). The material in model cwb4
does not quite reach the maximum speeds seen in model cwb1, most likely
because the combined radiative driving force, which is greater
the closer the stars are to each other, would have been weaker in 
model cwb4 prior to periastron. Model cwb4 also displays an excess of
mass relative to model cwb1 at speeds $<1400 \kmps$. 
Fig.~\ref{fig:statisticscwb4}(d) compares the distributions between
models cwb2 and cwb4 at apastron. They are remarkably similar,
except again model cwb4 displays a relative mass excess 
at speeds $<1400 \kmps$, which is likely due to the presence of
cold clumps. The distributions of hot ($T > 10^{5}\,$K)
gas are also remarkably similar.

Fig.~\ref{fig:statisticscwb4}(e) shows mass versus temperature
histograms from models cwb1 and cwb4 at periastron. Model cwb4 has
more than twice as much mass at temperatures above $10^{6}$\,K, but
less mass at lower temperatures, due to the reduced cooling in the
post-shock history of the downstream gas in model cwb4. At apastron,
model cwb4 has slightly less mass at $T > 8\times 10^{6}$\,K compared
to model cwb2 (Fig.~\ref{fig:statisticscwb4}f), but significantly more
mass between temperatures of $4-8\times 10^{6}$\,K.

The volume versus temperature histograms from models cwb1 and cwb4 at
periastron are shown in Fig.~\ref{fig:statisticscwb4}(g).  There is a
significantly greater volume of gas at $T\sim10^{6}$\,K in model
cwb4. This excess combines with a reduction in the relative volume of
gas at intermediate temperatures ($1.5 \times 10^{4} < T < 6
\times10^{5}$\,K). These differences reflect the hotter thermal
history of the shocked gas in model cwb4 prior to periastron. At
apastron in model cwb4, the volume of hot gas at temperatures above
$8\times10^{6}$\,K is $25-60$ per cent smaller than in model
cwb2. Model cwb4 generally also has smaller volumes of gas at
intermediate temperatures ($1.5 \times 10^{4} < T < 4 \times
10^{6}$\,K) -- this is likely a numerical effect due to the higher
resolution in model cwb4 compared to model cwb2, which leads to a
narrower layer of intermediate temperature cells at the unresolved
shocks. Model cwb4 has almost an order of magnitude greater volume of
gas at $5-6 \times 10^{6}$\,K than model cwb2. Such temperatures are
encountered in the mixing regions downstream of the clumps in model
cwb4 (c.f. Fig.~\ref{fig:cwb4_3dtemp2}).

\section{Summary and conclusions}
\label{sec:summary}
The model results described above reveal a wealth of interesting
dynamical effects which occur in short period, O+O binaries, including
radiative inhibition, rapid postshock cooling, and powerful
instabilities, all of which had been expected. However, there have
also been some surprising findings, such as the ability of cold dense
clumps to exist for substantial periods of time in the WCR when
current postshock conditions are largely adiabatic. Another surprise
is that the slow exit velocity of these clumps out of the system
allows some of them to exit the WCR through its trailing shock. More
generally, the presence of clumps increases the complexity of the WCR,
affecting the density, temperature and velocity of the shocked plasma,
and allowing additional material to be processed through normal
shocks. As a result, the emission from the WCR may be affected in a
non-trivial manner.

In future work we will study the multi-wavelength emission from these
models, and will repeat our study for WR+O systems where we expect
radiative braking to have a significant effect on the dynamics, and
the observed emission to show greater orbital modulation due to the
larger differences in the attenuation through the stellar winds.
Braking should also be important in O+O systems where there is a
larger ratio of wind opacities than explored here. We will also turn our
attention to modelling specific systems, with the aims of determining
key system parameters and understanding the natures of the diverse
range of wind-wind interactions which occur.

\section*{acknowledgements}
It is a pleasure to thank Ken Gayley, the referee, for a very helpful
report, and Stan Owocki and Ross Parkin for many interesting and
useful conversations.  I would also like to thank the Royal Society
for a University Research Fellowship.

\label{lastpage}

\end{document}